\documentclass[preprint]{article}

\usepackage{arxiv}

\usepackage[utf8]{inputenc} 
\usepackage[T1]{fontenc}    
\usepackage{hyperref}       
\usepackage{url}            
\usepackage{booktabs}       
\usepackage{amsfonts}       
\usepackage{microtype}      
\usepackage{lipsum}			
\usepackage{tikz}
\usetikzlibrary{shapes, arrows, positioning, fit, backgrounds}
\usepackage{graphicx}
\usepackage{subcaption}
\usepackage{amsmath}
\usepackage[linesnumbered,ruled]{algorithm2e}
\usepackage{graphbox}
\usepackage{color,soul}
\usepackage[affil-it]{authblk}

\usepackage{float}
\usepackage{setspace}   

\graphicspath{{../figures/}}

\title{Framework of Fracture Network Modeling using Conditioned Data with Sequential Gaussian Simulation}

\author[ ]{Yerlan Amanbek\textsuperscript{1,}\thanks{Corresponding author: yerlan.amanbek@nu.edu.kz}}
\author[ ]{Timur Merembayev\textsuperscript{1,2,}\thanks{Corresponding author: timur.merembayev@gmail.com}}

\author[3]{Sanjay Srinivasan}

\affil[1]{Department of Mathematics, Nazarbayev University, Kabanbay Batyr Avenue 53, Nur-Sultan, Kazakhstan}
\affil[2]{International Information Technology University, Almaty, Kazakhstan}
\affil[3]{Department of Energy and Mineral Engineering, Pennsylvania State University, 110 Hosler Building, University Park, State College, PA 16802-5000, USA  }

\affil[ ]{\textit{yerlan.amanbek@nu.edu.kz, timur.merembayev@gmail.com,  szs27@psu.edu}}

\renewcommand{\headeright}{}

\begin{document}
\maketitle

\large

\begin{abstract}

The fracture characterization using a geostatistical tool with conditioning data is a computationally efficient tool for subsurface flow and transport applications.The main objective of the paper is to propose a framework of geostatistical method to model the fracture network. In the method, we have chosen neighborhood area to apply the Gaussian Sequential Simulation in order to generate the fracture network in the unknown region. The angle was propagated from the seed where direction is guided by the neighborhood data in simulation regime. Initial seeds can be distributed by Poisson procedure. The method is applied for geological faults from the Central Kazakhstan and for field data from Scotland, UK. The simulation results are compatible with the original fracture network in the flow and transport modeling setting. From the research that has been carried out, it is possible to conclude that the numerical simulation of fracture network is a valuable tool in the subsurface flow and transport applications.

\end{abstract}

\keywords{fracture network model \and Sequential Gaussian Simulation \and geostatistical method}

\section{Introduction}
Fractures characterization plays vital role in flow and transport problems in the subsurface applications such as oil and gas production, groundwater remediation, CO2 sequestrations and etc. Due to limited information about the subsurface property, the prediction of the fracture network in the porous media is still a challenge. 

A variety of approaches has been studied to predict behavior of fracture at particular regions. For several years great effort has been devoted to numerical and experimental studies of fracture propagation described using complicated geomechanical equations using various methods \cite{mikelic2015phase, almani2017multirate, wick2014pressurized, shiozawa2019effect, shovkun2019fracture}. Such geomechanical models require more computational resources to model a fracture network with provided subsurface parameters. However, the key reasons for fracture propagation can be detected from such analysis. These reasons are necessary to adopt in the design of the fracture behavior. Input parameters have a large uncertainty due to spatial variations of subsurface properties such as stress field, existing fracture dimensions, fracture toughness, physical properties and etc. 

Permeability and porosity profiles are typically interpreted from the fracture geometric network, which can be generated stochastically. This structure might not be exactly resemble a real fracture structure, however, it provides a reasonable configuration for the flow and transport models. On the basis of the available data, for example data collected near wellbore and other techniques, a geostatistical technique can be trained based on known regions to model the fracture network in unknown regions.  In the literature, stochastic approaches have attracted much attention from researchers in reduction of subsurface uncertainty. 

In \cite{hyman2015dfnworks} the dfnWorks workflow was introduced to simulate single or multi-phase flow and transport problem coupled with geostatistically generated discrete fracture network (DFN). A combination of different types of tools in single platform is necessary to validate modeling uncertainty.

In \cite{liu2002geological} the geostatistical modeling of fracture system was developed using pattern statistics from  pattern bar chart. The multiple point statistics(MPS) was used with pixel based data in the growth-based fracture network method. As reported in \cite{chugunova2017explicit}, the model was trained from satellite images which were obtained by detecting faults in the surface. The training and testing were performed for chosen surface samples.  

Authors \cite{jung2013training} have proposed the fracture network system simulation based on MPS \cite{chugunova2017explicit,chandna2019modeling} using the sophisticated relations among cracks and faults. In \cite{chandna2019modeling} the method was developed based on MPS honoring surrounding data and multiple point histogram. The propagation orientation is utilized  in the MPS simulation. This method was applied for real-world datasets such as Teapot Dome, Wyoming. Another application of discrete fracture network simulation work using a commercial software MoFrac were presented in \cite{junkin2018analysis} for the Canadian Shield dataset.

An interesting approach to this issue has been proposed by \cite{srivastava2005geostatistical}.  This method is adaptable for additional user defined constraints which honors other types of subsurface data. Algorithms including Sequential Gaussian Simulation (SGS) is used to model with simple kriging approach of the fracture orientation. In the model, the process is similar to a propagation of fracture where in each iteration the orientation and step length is computed honoring provided subsurface information. However, there is a necessity on systematic workflow in the fracture pattern prediction.

In this paper, we present the fracture characterization model using geostatistical analysis with  conditioned data based on the approach in \cite{srivastava2005geostatistical}. The workflow of algorithm is provided for the natural fracture network simulation in the undiscovered zone. Numerical simulations were conducted for different examples such as realistic faults from the Central Kazakhstan and the natural fracture dataset from South Glasgow, Scotland, UK. To verify the realization, we have compared major variables such as the orientation histogram and the concentration profile at production well, where the adaptive numerical homogenization method was applied to simulate the single-phase and incompressible flow and transport model in the fractured reservoir due to multiscale feature in this setting, see \cite{amanbek2019adaptive, singh2017adaptive} for more details.    

The remaining part of the paper proceeds as follows. Section \ref{sec:num_meth} describes the numerical method. The process of modeling is shown in the flow-chart. Numerical results are presented in Section \ref{sec:num_res}. In Section \ref{sec:disc}, the results were discussed and the conclusion is reported in Section \ref{sec:concl}.

\section{Numerical Methods}
\label{sec:num_meth}

We first illustrate the fracture network simulation in 2D with neighbor selection procedure, and then the Sequential Gaussian Simulation (SGS) process for simulating fracture direction. In our case, there is data from certain regions is provided, however, the rest of the region is unknown due to sparsity of data. By using the information of fracture network we model fractures in unknown regions. In the simulation of fracture networks, key parameters are fracture direction (angles) and length of propagation. The orientation or the angle of fracture was computed by SGS method using the data from known regions. In each iteration of simulation, fracture grows in all domain honoring the information of neighbourhood region.

It is common to observe the family of fractures, not only single  fracture. Such family of fractures has a common features including orientations. The changes of directions in the fracture family are usually smooth and transition of certain patterns are soft \cite{gringarten1999geometric}. Therefore, the average value of fracture orientation, the intensity distribution, etc are necessary in the fracture simulation.


\subsection{Fracture Network Simulation with Conditioned by Neighboring Data}
\label{sec:neigh}

From a geomechanical standpoint, a fracture direction is influenced the systematic organization of neighboring subsurface property such as geomechanical features, fracture intersections and others \cite{gringarten1999geometric}. The stress field nearby the fracture tip is a key factor in  the simulation of fracture propagation. This is measured by a stress intensity factor ($K_I$), which depends on loading (stress) and the fracture dimensions. The fracture keeps its propagation as long as $K_I > K_{IC}$, where $K_{IC}$ is fracture toughness. In addition, previous studies indicate that the energy minimization is important in the systematic fracture path simulation using the phase-field model \cite{mikelic2015phase, wick2014pressurized}. Therefore, the effect of neighbor fracture should be taken into account fracture direction simulations. 

We describe the key ingredients of algorithm in the fracture network simulations including  initiation, propagation and termination \cite{gringarten1999geometric}. We will address all steps below.

To illustrate the proposed method of fracture characterization, we show the major steps  in 2D for surface traces.

\begin{figure}[H]
  \centering
  \includegraphics[scale=0.16]{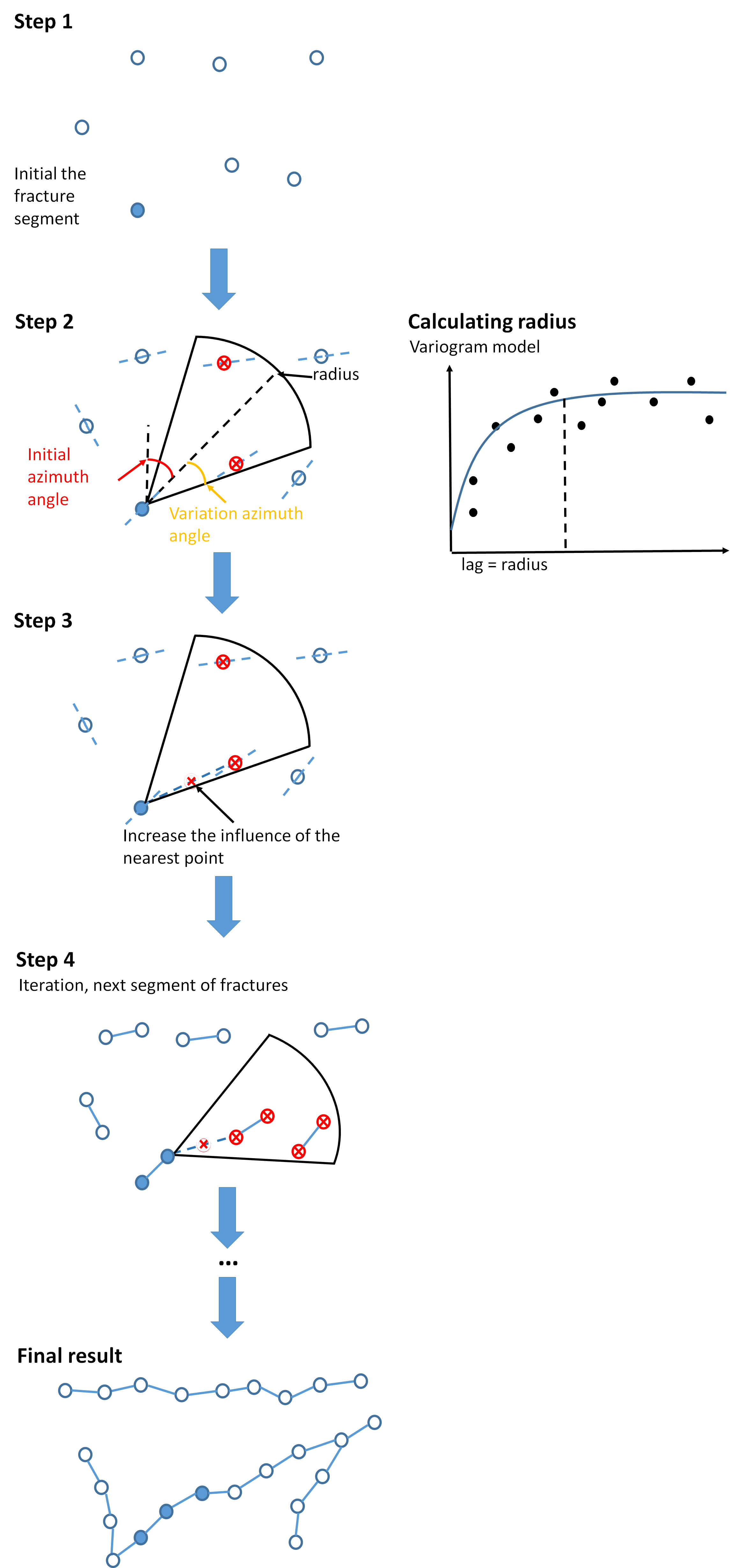}
  \caption{Schema of a fracture simulation.}
  \label{fig:fig1}
\end{figure}

In our model, a fracture is a collection of fracture segments with corresponding length and orientation. The initial fracture segment will be growing honoring the surrounding information. We first define the initial fracture segments (coordinates and azimuth angles) in the entire domain. In the unknown region, we use the Poisson procedure to seed initial fracture segments, however, these locations can be indicated by user. In known area, fractures will be propagated from the middle of the original fracture length and repeat their original path in both directions from the middle. Below you can see the key steps in the method for defining the neighborhood region, as shown in  Figure \ref{fig:fig1}.
\begin{itemize}
\item [Step 1.] The initial points will be seeded in the domain. For each point, which is the middle of fracture segment, there are corresponding parameters such as azimuth angles, coordinates of points and the neighboring segments. The filled point is target point from where a new fracture segment will grow.  
\item [Step 2.] We define the neighborhood zone as sector of influencing points that was created from the following parameters: the initial azimuth angle, a standard deviation of angle and radius. From the initial point along the line, which follows provided azimuth, we draw sector with radius r and angle of sector equals to two standard deviations. The line generated from azimuth divides the sector angle equally. In here, we compute azimuth of angle from fracture angle distribution of known region. A radius of the sector is defined from the variogram model of fracture segments in the known region. It is within this search sector that conditioning data for fracture angles will be searched. 
\item [Step 3.] There will be included a virtual point additionally to the selected points in the sector. To consider the influence of the nearest fracture segment, we draw virtual fracture segment in the center of the segment, which connects points between the current point and the nearest fracture segment. The virtual fracture segment keeps the angle. The selected points in the sector including the additional data will be used in the SGS. This approach was initially proposed in \cite{srivastava2005geostatistical}.
\item [Step 4.] We compute important features of a new fracture segment such as the orientation and the fracture segment length. We use the points in the neighbourhood sector as input data in the SGS analysis to compute a newly simulated angle. The angle is chosen from the computed distribution in random fashion. The length of  simulated segment equals the mean of fracture segment distribution of known region. We note that the length can be specified by user. Combining the information of angle and length of segments we draw a new fracture segment which is colored in blue.
\end{itemize}
By propagating the target or initial points and repeating step 2 - step 4 we can achieve a configuration shown in Figure \ref{fig:fig1} (Final result). Standard deviation and radius are constant for further iterations.  
After several propagation iterations  a newly simulated fracture will be terminated until crossing any fractures or length of fracture reaches the constraints. 
\newpage
\subsection{Sequential Gaussian Simulation}
\label{sec:SGS}
We briefly go through SGS algorithm in perspective  of fracture segment orientation.  In step 4 from the previous section, we need to identify  a new angle of the next fracture segment  by considering angles of neighboring fractures.

The method of SGS assumes the presence of a normal distribution of the simulated random variable. For any point in the region, the local distribution function will be distributed according to the normal law and will be determined by two parameters  such as mean and standard deviation.
Before to make simulation, it is assumed that the stationary is exist for random function $Z(x)$ and a random function $Y(x)$ is exist such that
\begin{equation}
Y(x)=\phi[Z(x)]
\end{equation}
where $Y(x)$ is $N(0,1)$  and $\phi$ - the normal score of the transformation. 

The ordinary kriging is the main method that helps to propagate the fracture  segment with the provided spatial distribution. The kriging is the linear unbiased predictor and can be explicit as the linearly - weighted average function of observations in the unknown area of fractures $x$, location of fractures . $Y(x)$ is a set of random variables (azimuthal fault angles). The ordinary kriging estimate $Y_{sk}^{*}(x)$ at $x$ is calculated as
\begin{equation}
Y_{sk}^{*}(x)=\sum_{i=1}^{n(x)} \lambda_{i}^{sk}(x)Y(x_i) 
\end{equation}

The ordinary kriging weights $\lambda_{i}^{sk}$ that defined by by solving the system $n(x)$ of ordinary kriging equations:
\begin{equation}
\sum_{j=1}^{n(x)} \lambda_{j}^{sk}C_{ij}=C_{i0}, \forall i=1,...,n(x)
\end{equation}
where $n(x)$ is the total number of neighboring points $x_i$ used in estimating the point $x$.
The ordinary kriging score variation is given by
\begin{equation}
\sigma_{sk}^{2}=C(0)- \sum_{i=1}^{n(x)} \lambda_{i}^{sk}C(x_i-x)
\end{equation}
As a result, we obtain the parameters of the local normal distribution function $N(Y_{sk}^{*}(x),\sigma_{sk}^{2}(x))$ at the estimation point $x$. This distribution is required in the identification of the fracture propagation angle. 

\subsection{Estimation of fracture orientation}
\label{sec:estim}
In step 4, we have used the SGS for the modeling of fracture orientation  in each iteration of the fracture network simulation. The SGS is the most commonly used sequential simulation algorithm for modeling continuous variables. A detailed introduction to this method can be found in \cite{goovaerts1997geostatistics, remy2009applied}. For visual representation of the flow-chart of the proposed method the reader is referred to Figure \ref{fig:fig2}.  In addition, the pseudocode of the proposed method is depicted in Figure \ref{fig:fig3}. In the algorithm,simulation of azimuth angles in SGS proceeds as follows:
\begin{enumerate}
  	\item [1.] Validate whether the distribution of the provided areas dataset is normal distribution or not. This dataset is a collection of the fracture segment azimuth; 
  	\item [2.] If necessary, transform the distribution of the data into a normal distribution;
	\item [3.] Seed  the initial fracture segments randomly using the Poisson process in unknown region;
	\item [4.] Identify the points in the neighbor zone as described in Step 2-3, Section \ref{sec:neigh};
	\item [5 \& 6.] If number of neighbors are less or equal 1 then the direction of the new fracture segment will be chosen by the Monte-Carlo method for the normal distribution of the provided dataset, as described in Step 1, Section \ref{sec:neigh}. This produces a new azimuth angle that is combined with the length of fracture segment in order to propagate a new fracture segment; 
	\item [5 \& 7.] If number of the neighbor points are more than 1 then to add point of halway to the nearest neighbor for the influence of the nearest fracture segment;
	\item [8 \& 9.] As subroutine of SGS procedure, the  ordinary kriging approach is utilized to obtain kriged estimate and the corresponding kriging variance;
	\item [10.] Termination of the fracture segment occurs when it crosses another one. We do not consider another option for termination a propagation of fractures, criteria of achievement limit length of full length of fracture, the criteria can be integrated;
	\item [11.] Transform from the normal distribution to the original univariate distribution.
\end{enumerate}
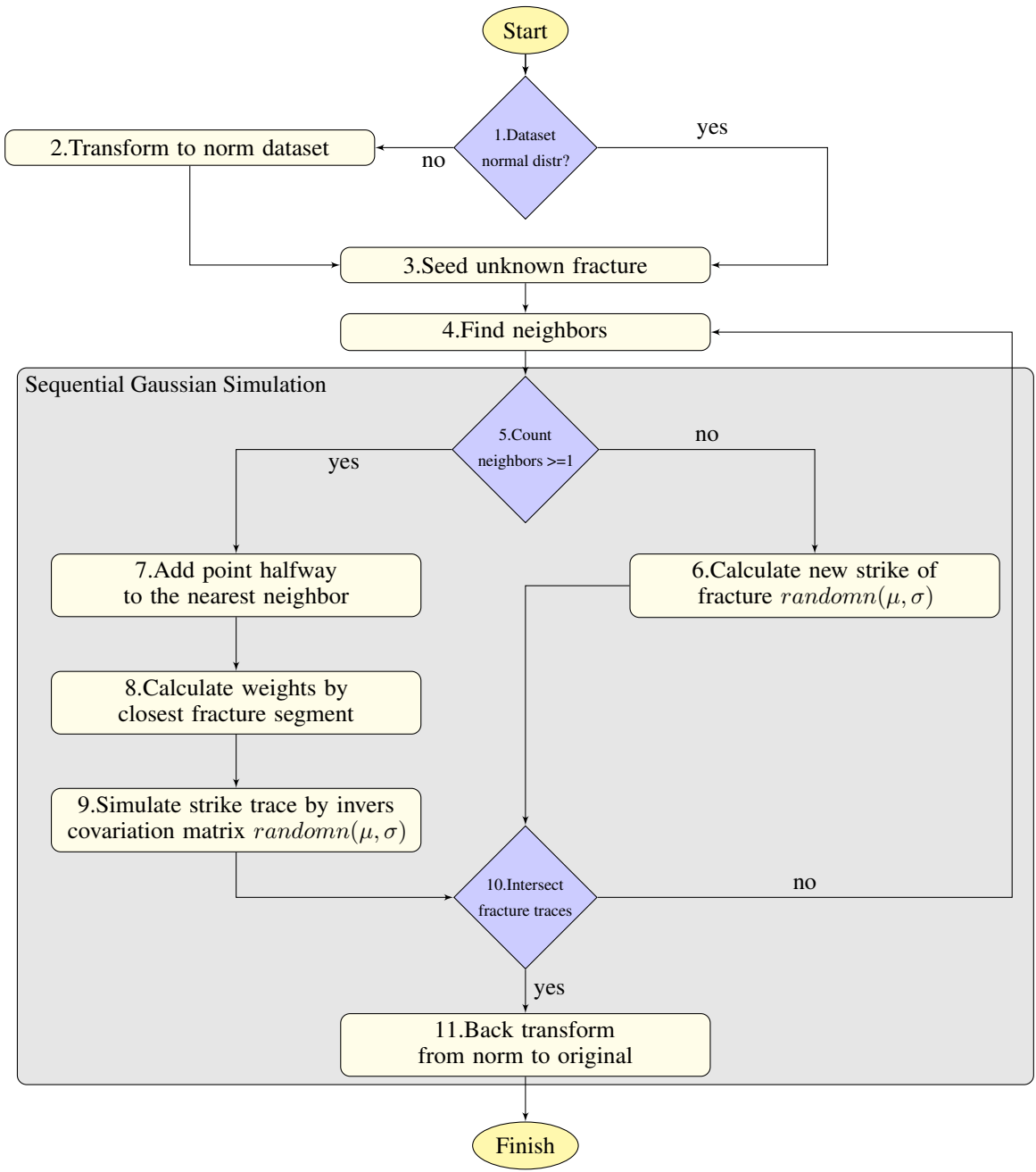
\begin{figure}[H]
  \centering
	\tikzstyle{decision} = [diamond, draw, fill=yellow!40, 
    	text width=4.5em, text badly centered, node distance=1.75cm, inner sep=0pt]
	\tikzstyle{decisiong} = [diamond, draw, fill=blue!20, 
    	text width=4.5em, text badly centered, node distance=1.75cm, inner sep=0pt]

	\tikzstyle{block} = [rectangle, node distance=1.75cm, minimum width=1cm, minimum height=0.5cm, 			draw, fill=yellow!10, text width=15em, text centered, rounded corners, minimum 			height=1.5em]
	\tikzstyle{blockg} = [rectangle, minimum width=1cm, minimum height=0.5cm, draw, fill=blue!10, 
    	text width=15em, text centered, rounded corners, minimum height=1.5em]
	\tikzstyle{line} = [draw, -latex']
	\tikzstyle{cloud} = [draw, ellipse,fill=yellow!40, 
		node distance=1.5cm, 
    	minimum height=2em]
	\tikzstyle{invis} = [draw, fill=yellow!10, 
		node distance=4.25cm, 
    	minimum height=2em]
	\tikzstyle{invisg} = [draw, fill=yellow!10, 
		node distance=3.5cm, 
    	minimum height=2em]
	\tikzstyle{matheq} = [node distance=8.75cm, text width=21em, minimum width=1cm, 
    	minimum height=2em, text centered]
    \tikzstyle{blockjusttext} = [rectangle, node distance=1.75cm, minimum width=1cm, minimum height=0.5cm,
    	text width=20em, text centered, rounded corners, minimum height=1.5em]
	\tikzstyle{blockfill} = [rectangle, node distance=1.75cm, minimum width=1cm, minimum height=14cm, draw, fill=yellow!10, 
    	text width=20em, text centered, rounded corners, minimum height=16.8em]

	\begin{tikzpicture}[ node distance = 1.25cm, auto]
    	\node [cloud] (init) {Start}; 
    	\node [decisiong, below of=init] (conver) {{\scriptsize 1.Dataset normal distr?}};
    	\node [block,left of=conver, node distance=5cm] (second) {2.Transform to norm dataset};
    	\node [block,below of=conver, node distance=1.75cm] (third) {3.Seed unknown fracture};
    	\node [block, below of=third, node distance=1cm] (fourth) {4.Find neighbors };
    	\node [decisiong, below of=fourth] (conver1) {{\scriptsize 5.Count neighbors >=1}};
    	\node [block, below left=1cm and 1cm of conver1] (fifth) {7.Add point halfway to the nearest neighbor};
    	\node [block, below right=1cm and 1cm of conver1] (sixth) {6.Calculate new strike of fracture $randomn(\mu,\sigma)$};	
		\node [block, below of=fifth] (seventh) {8.Calculate weights by closest fracture segment};
		\node [block, below of=seventh] (eighth) {9.Simulate strike trace by invers covariation matrix $randomn(\mu,\sigma)$};
		\node [decisiong, below = 4.5cm of conver1] (conver2) {{\scriptsize 10.Intersect fracture traces}};
		\node [block, below of=conver2, node distance=2.2cm] (ninth) {11.Back transform from norm to original};
		\node [cloud, below of=ninth, node distance=1.5cm] (finish) {Finish};
    	\begin{scope}[on background layer]
    		\node[draw,fill=black!10, fit=(conver1)(eighth)(sixth)(ninth), inner xsep=5mm, rounded corners] (backg){};
    		\node[below right] at (backg.north west) {Sequential Gaussian Simulation};
    	\end{scope}
    
    	\path [line] (init) -- (conver);
   		\path [line] (second)|- (third);
   		\path [line] (third) -- (fourth);
   	 		\coordinate[right of=third] (a1);  
    		\coordinate[left of=conver] (b1);  
    		\coordinate[right of=conver] (e1);
    		\coordinate[left of=second] (f1);
    	\path [line] (init) -- (conver);
    	\path [line] (conver) -- node [near start] {no} (second);
    	\path [line] (conver) -| node [near start] {yes} ([xshift=3.25cm]e1) --([xshift=3.25cm]a1)-- (third);
		\path [line] (fourth) -- (conver1);
    	\path [line] (conver1) -| node [near start] {no} (sixth);
    	\path [line] (conver1) -| node [near start] {yes} (fifth);
		\path [line] (fifth) -- (seventh);
   		\path [line] (seventh) -- (eighth);
		\path [line] (eighth) |- (conver2);
		\path [line] (sixth) -| (conver2);
			\coordinate[right of=fourth] (a3);  
    		\coordinate[left of=conver2] (b3);  
   			\coordinate[right of=conver2] (e3);
    		\coordinate[left of=ninth] (f3);
    	\path [line] (conver2) -| node [near start] {no}([xshift=6cm]e3) --([xshift=6cm]a3)-- (fourth);
    	\path [line] (conver2) -- node {yes} (ninth);
		\path [line] (ninth) -- (finish);

	\end{tikzpicture}
	\caption{Flow-chart of the fracture algorithm.}
	\label{fig:fig2}
\end{figure}

\begin{figure}[H]
  \centering

	\begin{algorithm}[H]
    	\SetKwInOut{Input}{Input}
    	\SetKwInOut{Output}{Output}

    	\Input{mean, standard deviation $std$, fracture length $len$, iteration $iter$, matrix azimuth $A$, count of seed fractures $m$}
    	\Output{$A(m,n)$}
    	\For{$i \gets 1$  \KwTo $m$} 
    		{
    		\For{$j \gets 1$  \KwTo $iter$} 
    			{
    			$[x_{id_{ij}},y_{id_{ij}}] \gets $ \textit{Calculate of a mid of fracture}\\
    			$[x_{id_k},y_{id_k}]  \gets $ \textit{Find neighbours $A(X,Y)$}\\
    			\eIf{$count[x_{id_k},y_{id_k}]\leq 1$}
    				{
    				$x_{id_{i,j+1}}=len*cos(randomn(mean,std))+x_{id_{ij}}$\\
    				$y_{id_{i,j+1}}=len*cos(randomn(mean,std))+y_{id_{ij}}$\\
    		
    				\If{$A(x_{id_{ij}},y_{id_{ij}}) \cap A(X,Y)$}
      					{
      					\textit{close loop j}
    					}
    				}
    				{
    				$\lambda,\sigma \gets $ \textit{Calculate weights($id_k, covar^{-1}$)}\\
    				$A(x_{id_{i,j+1}},y_{id_{i,j+1}})=$\textit{Simulation($\lambda,\sigma,randomn(mean,std)$)}\\
    				\If{$A(x_{id_{ij}},y_{id_{ij}}) \cap A(X,Y)$}
      					{
      					\textit{close loop j}
    					}
    				}
    			}
    		}
    	\caption{Sequential Gaussian Simulation}
	\end{algorithm}
  \caption{Pseudo code of fracture characterization .}
  \label{fig:fig3}
\end{figure}

\section{Numerical results}
\label{sec:num_res}

In this section, we show several numerical examples such as comparison with previous result from \cite{liu2002geological}, application for faults from Zhezkazgan, Kazakhstan and fractures of McDonald limestone dataset from Scotland, UK. To illustrate applications of the proposed method, we show numerical simulations for realistic datasets  in two dimensions with verifications.

\subsection{Example 1}
\label{sec:expl1}

\begin{figure}[H]
  \begin{subfigure}{.5\textwidth}
  		\centering
  		\includegraphics[width=.8\linewidth]{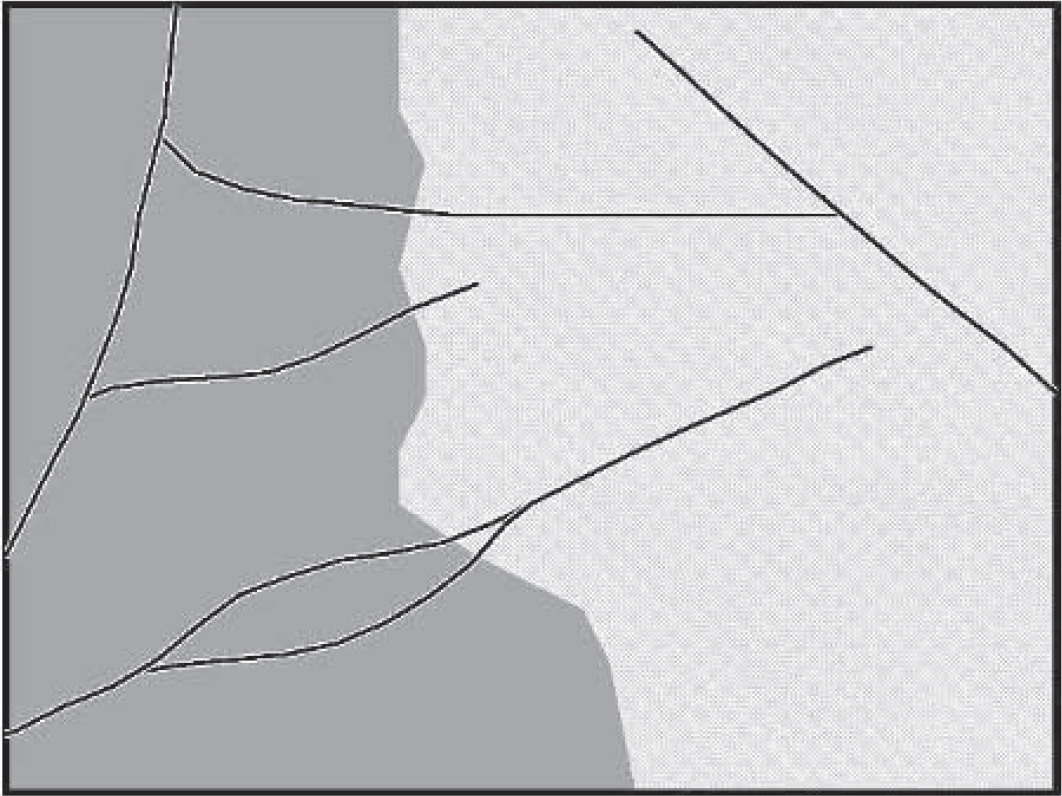}
  		\caption{The result of simulation in \cite{srivastava2005geostatistical}.}
  		\label{fig:fig4a}
  	\end{subfigure}
  	\begin{subfigure}{.5\textwidth}
  		\centering
  		\includegraphics[width=.85\linewidth]{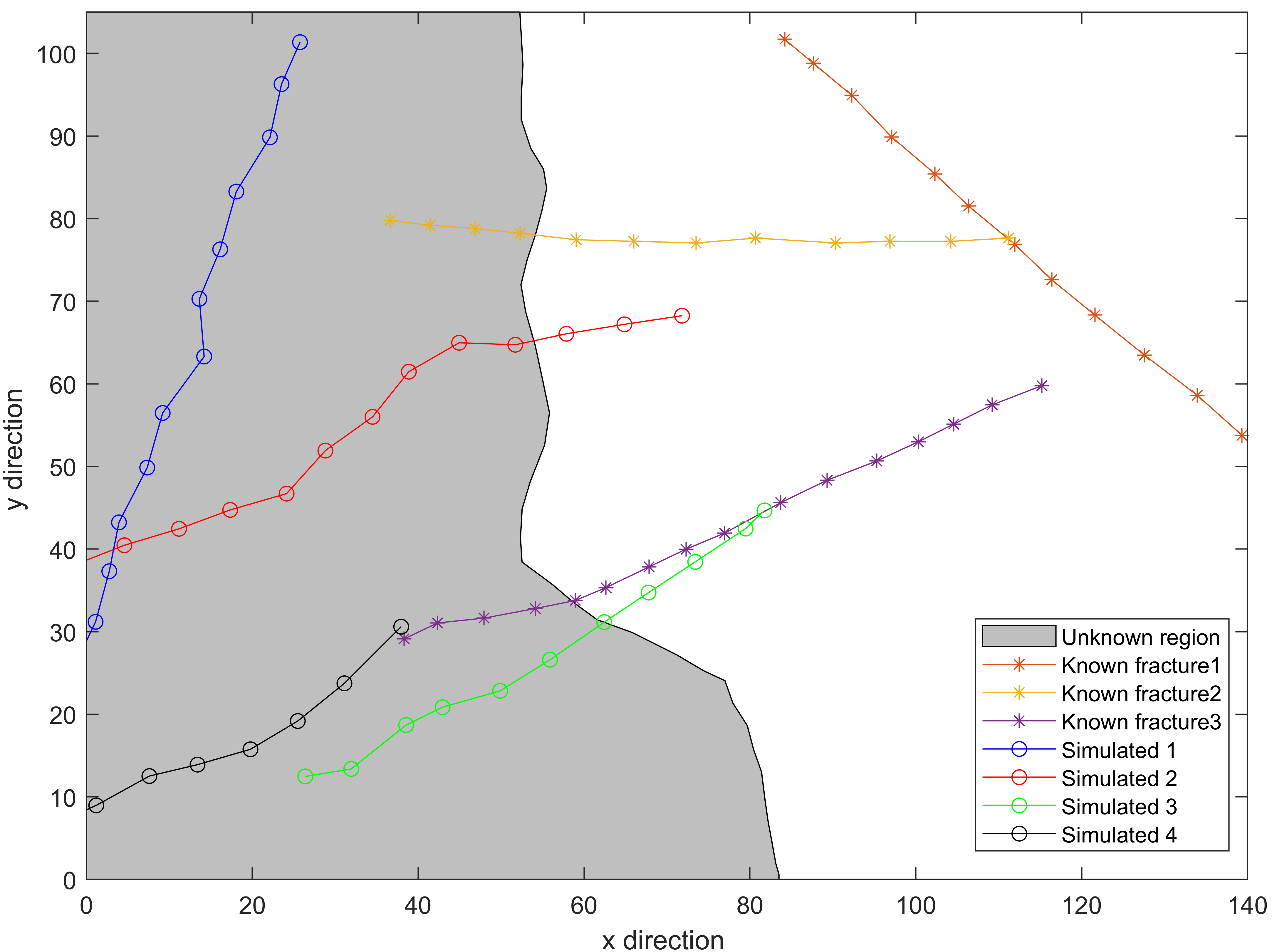}
  		\caption{Our result of simulation.}
  		\label{fig:fig4b}
  	\end{subfigure}
  \caption{The results of simulation by proposed algorithm.}
  \label{fig:fig4}
\end{figure}

In this example, we have applied our method to the scenario with initial conditions presented in \cite{srivastava2005geostatistical}.   We have compared the results in here to depict the similarity of proposed method. To verify the method, our result of simulation is compared with the result of fractures from \cite{srivastava2005geostatistical}, as shown in Figure \ref{fig:fig4}.

The example was digitized by using GRABIT \cite{doke2005grabit}. Fractures from the gray region is given in the right side. In the original work, fracture of the dark region was predicted. We have used the following parameters as input data in our algorithm. The mean of orientations is 70 degrees and the standard deviation is 12. The variogram model is a spherical. Nugget is 1, spherical is 2, range is 50, r = 50, fracture length is 10. 
  
In here, the azimuth angles were generated by Monte Carlo method in the SGS procedure. As can be seen from Figure \ref{fig:fig4b}, simulated fractures are presented in different colors such as green, blue, red and black. It is  important to note that the fracture path in the known region was reproduced and the part of it was honored during the simulation regime. 

\begin{figure}[H]
	\begin{subfigure}{.5\textwidth}
		\centering
		\includegraphics[width=.8\linewidth]{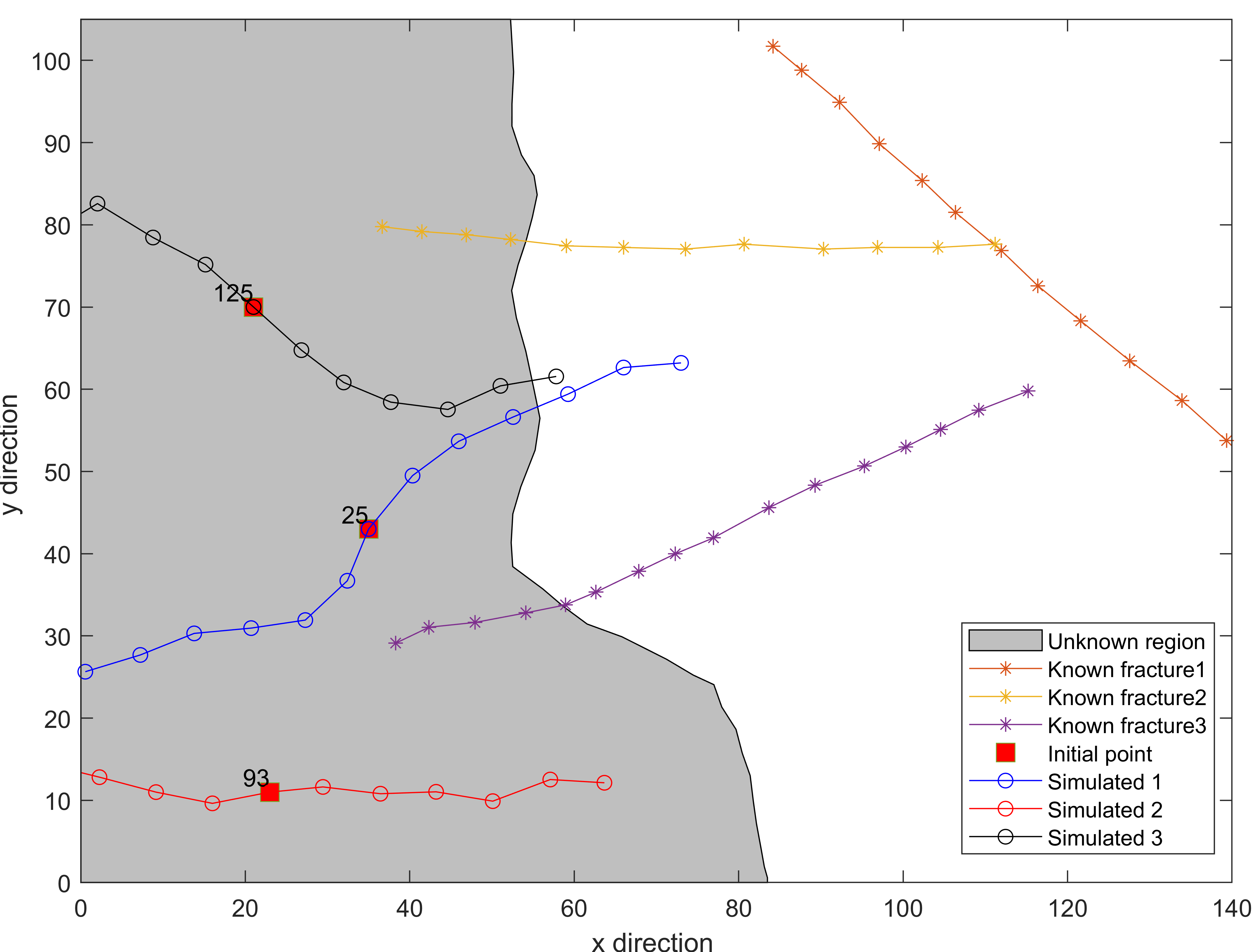}
		\caption{Example 1.}
		\label{fig:fig5a}
	\end{subfigure}
	\begin{subfigure}{.5\textwidth}
		\centering
		\includegraphics[width=.8\linewidth]{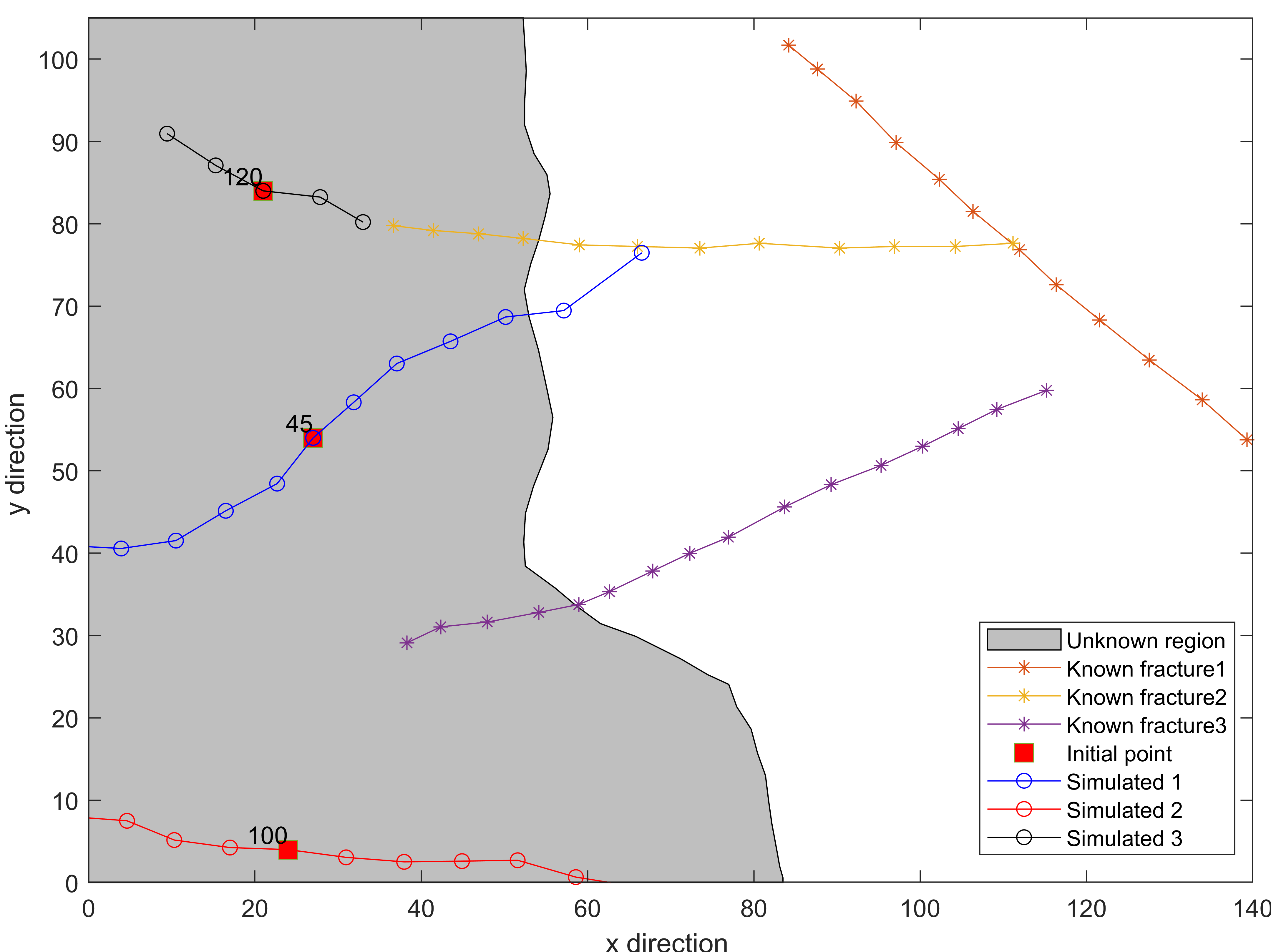}
		\caption{Example 2.}
		\label{fig:fig5b}
	\end{subfigure}
	\caption{Simulation fractures from initial points generated by Poisson process.}
	\label{fig:fig5}
\end{figure}

Various numerical examples are performed with different initial seeds in the hidden region.  Poisson procedure is applied to determine coordinates of initial points for simulated fractures. 
Figure \ref{fig:fig5} presents different realizations where red squares are middles of initial seeds.
In Figure \ref{fig:fig5a} and \ref{fig:fig5b}, blue fractures (initial azimuth: 45 and 25) are similar to the original fracture. During the simulation, we have observed that there is a tendency of having traces in similar directions for nearby fractures. This phenomena was discussed in Section 2.1 with perspective of geomechanics. We observe similar pattern for a black fracture (initial azimuth 120), which connected to the known fracture, see Figure \ref{fig:fig5a}, Figure \ref{fig:fig5b} gives an examples of similar direction pattern for a black and blue fractures(initial azimuth 125 and 25). After several more iterations the resulting path of black and blue fractures become close to each other in the center of domain.

\subsection{Example 2}
\label{sec:expl2}

The analysis of conditioned data is necessary for the fracture characterization simulation. We first provide the results obtained from the FracPaq toolbox in Matlab \cite{healy2017fracpaq}. Second, we show the one of realizations of simulated fractures for the region in Zhezkazgan, Kazakhstan.

\subsubsection{Fracture Data analysis}
\label{sec:fracan}

Before applying SGS algorithm we need to make data analysis of fractures given in known area using FracPaq \cite{healy2017fracpaq}. The software allows to obtain the quantitative fracture patterns, digitalization of fractures and their distribution in two dimensions for length, segment scales, and rock types. Fracture is defined as a collection of one or more segments which creates a continuous line in the software. It is assumed that segments at any scale are formed because of the interaction and fusion of small cracks, so analysis of these segments separately can be useful.

\begin{figure}[H]
	\begin{subfigure}{.5\textwidth}
		\centering
		\includegraphics[width=.8\linewidth]{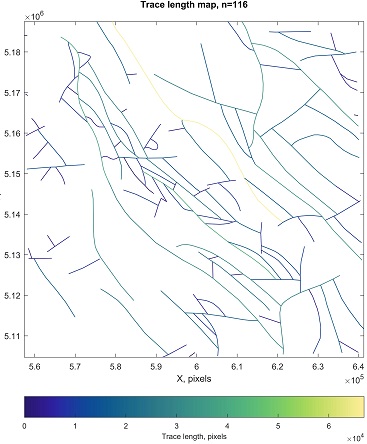}
		\caption{Traces colored by length.}
	\end{subfigure}
	\begin{subfigure}{.5\textwidth}
		\centering
		\includegraphics[width=.8\linewidth]{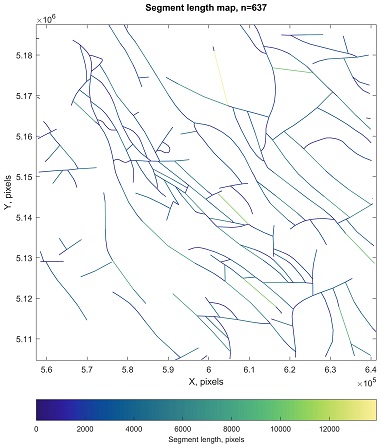}
		\caption{Segments colored by length.}
	\end{subfigure}
	\caption{Result of data analysis.}
	\label{fig:fig6}
\end{figure}

The FracPaQ provides an analysis of the traces and the segment of fractures by length so it helps to indicate major and minor fractures, analysis of segment traces provides visual confirmation that fractures have one mean of fracture segments. The traces colored and the segment colored of fractures by length are shown on Figure \ref{fig:fig6}.

\begin{figure}[H]
  \begin{subfigure}{.5\textwidth}
  		\centering
  		\includegraphics[width=.8\linewidth]{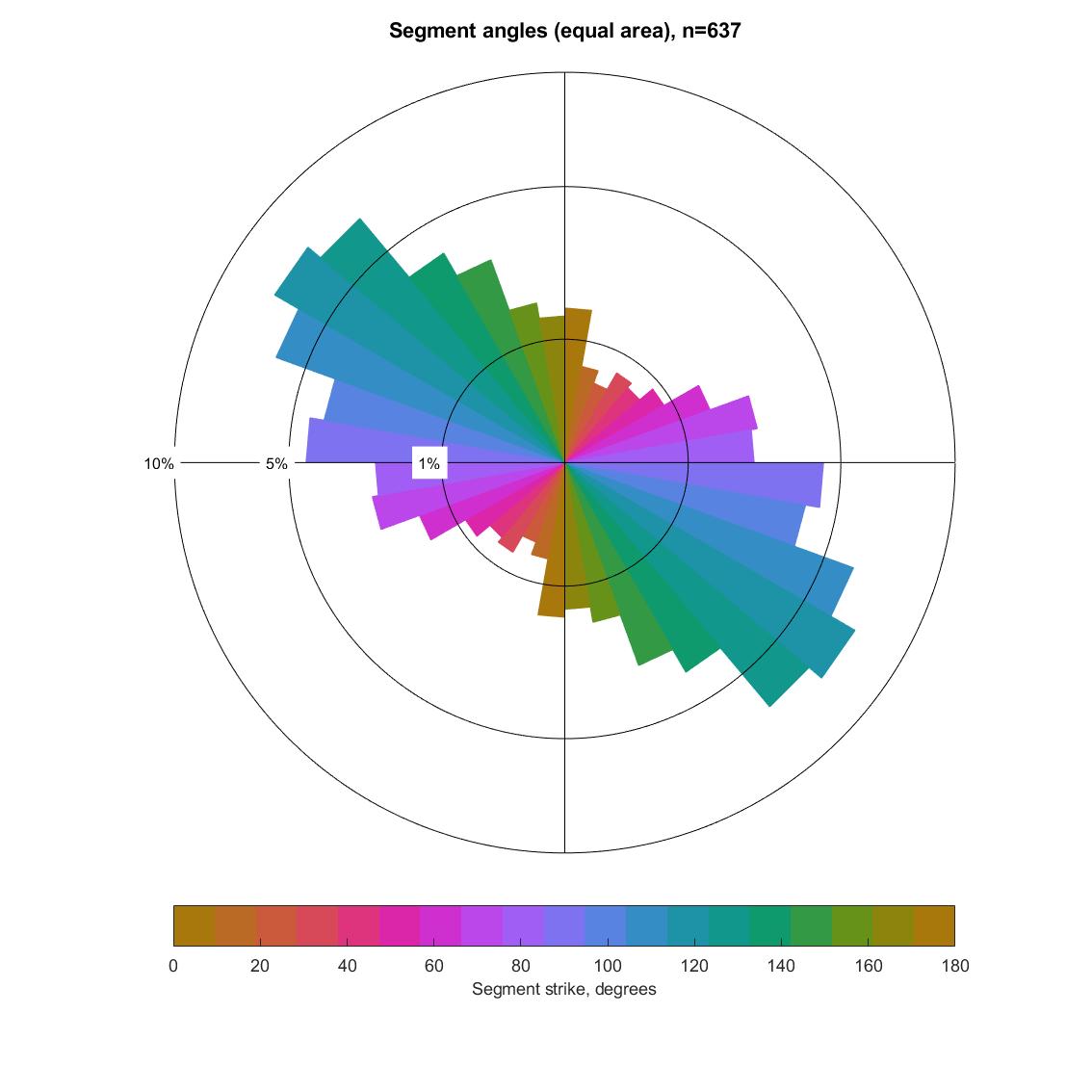}
  		\caption{Angles of fracture segments.}
  	\end{subfigure}
  	\begin{subfigure}{.5\textwidth}
  		\centering
  		\includegraphics[width=.8\linewidth]{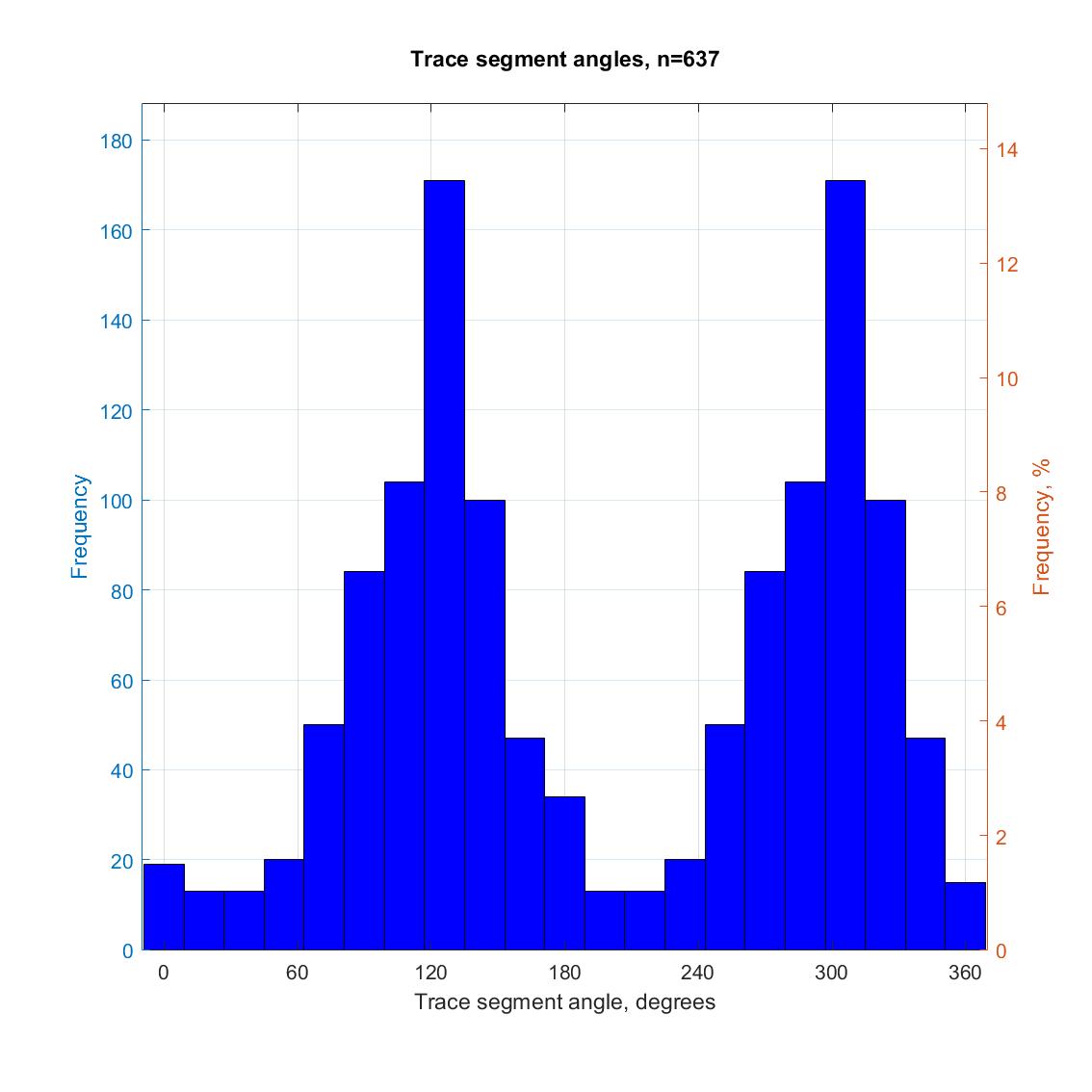}
  		\caption{Angles of fracture segments.}
  	\end{subfigure}
  \caption{Fracture orientation.}
  \label{fig:fig7}
\end{figure}

\begin{figure}[H]
  \begin{subfigure}{.5\textwidth}
  		\centering
  		\includegraphics[width=.8\linewidth]{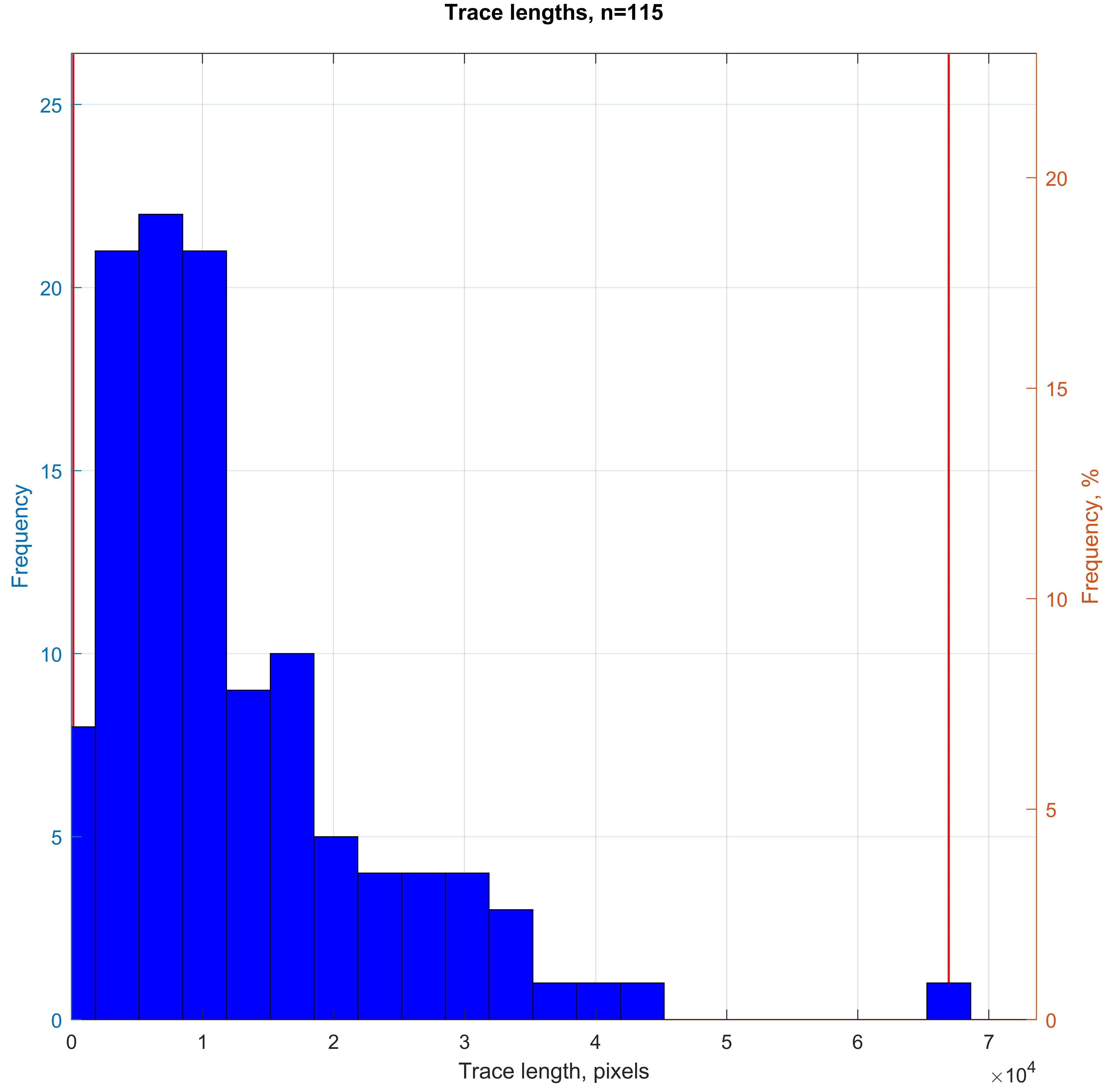}
  		\caption{Frequency-size histogram for trace lengths.}
  	\end{subfigure}
  	\begin{subfigure}{.5\textwidth}
  		\centering
  		\includegraphics[width=.8\linewidth]{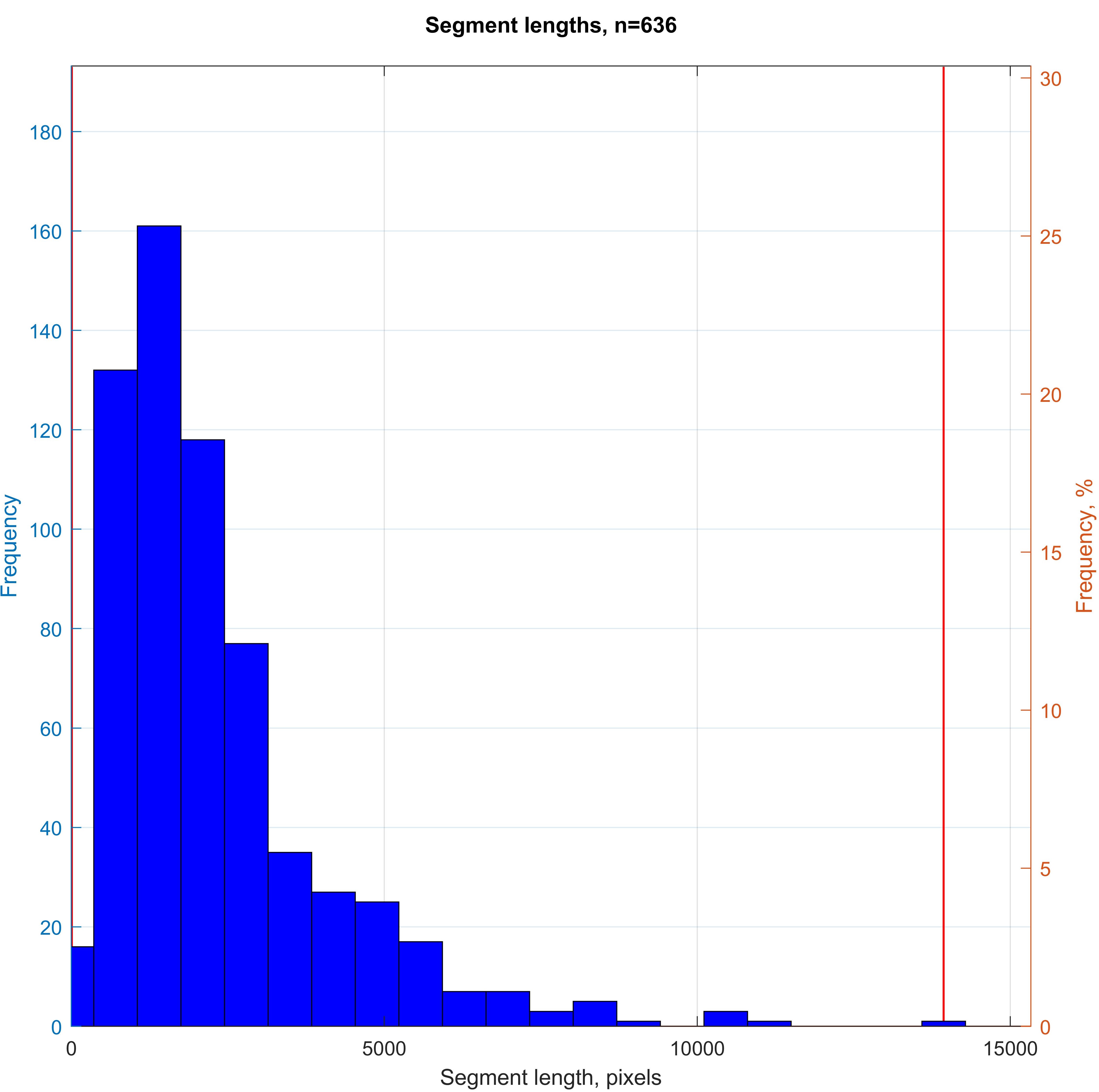}
  		\caption{Frequency-size histogram for segment lengths.}
  	\end{subfigure}
  \caption{Histograms of analyzed data. Red lines show the minimum and maximum lengths.}
  \label{fig:fig8}
\end{figure}

The distribution of fracture angles for the known region is shown in Figure \ref{fig:fig7} and is presented in a rose plot and a histogram. The rose plots and the histograms show the main fracture population (about 120 degrees) in the known region, the second peak of the fracture population is an opposite direction (120 degrees) a fracture segment. Based on the angles shown in the rose plot, the mean of the azimuth angle of fracture segments is 120.

Base on the data there are two frequency distributions  – size plots for trace and segment lengths can help to estimate the mean length of segment. The length will be used for simulation of fracture propagation. Figure \ref{fig:fig8} shows histograms of trace and segments lengths of geology faults.

FracPaQ generated  the variogram (or semi-variogram) of segment lengths. Semivariogram had been calculated as a function of the separation of two segments from the stored coordinate positions of all the segment mid-points (centroids) for every pair of segments. 

\begin{figure}[H]
  \centering
  \includegraphics[width=.6\linewidth]{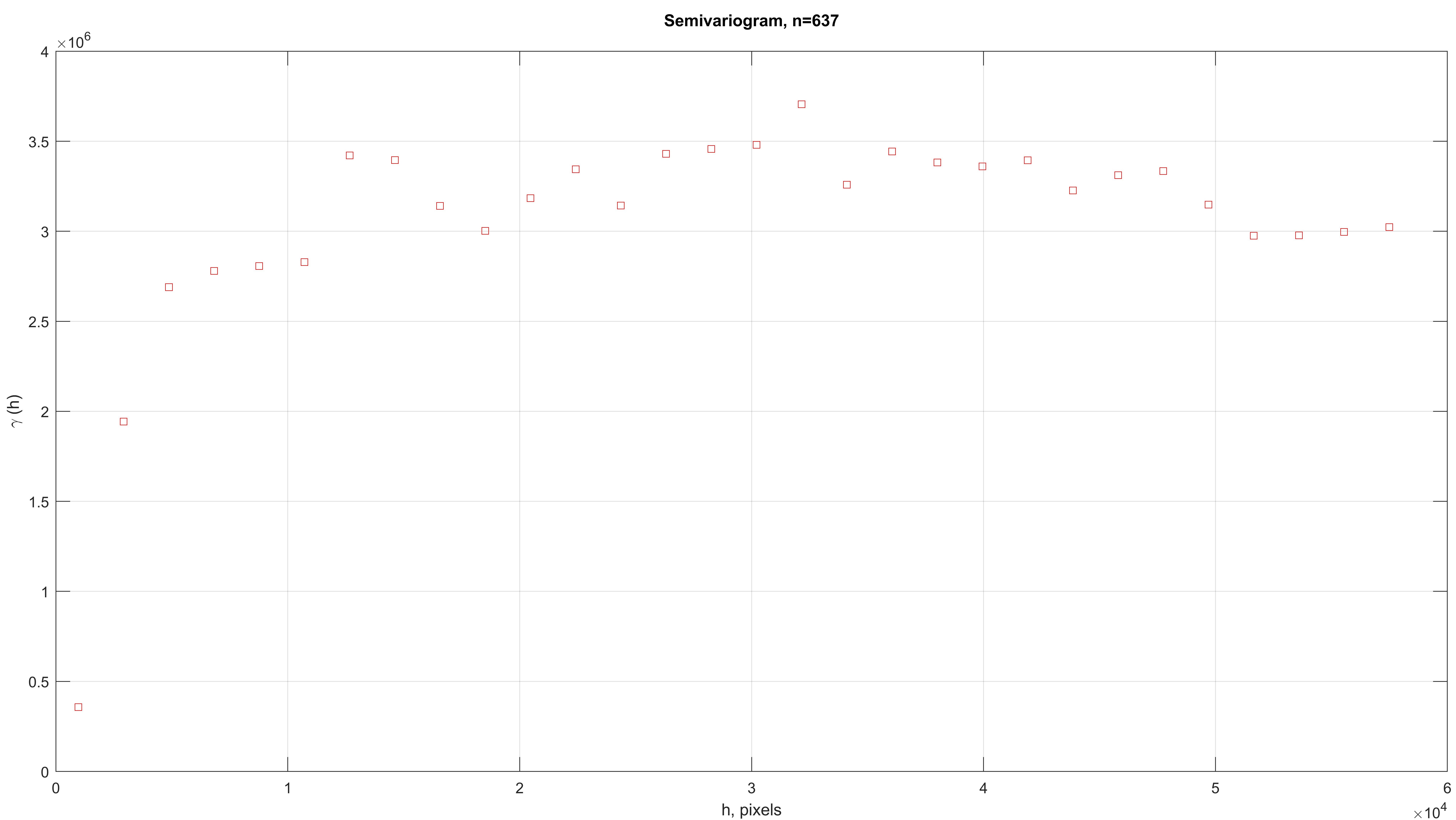}
  \caption{Variogram of segment lengths.}
  \label{fig:fig9}
\end{figure}

In Figure \ref{fig:fig9} the calculated variogram model showed the correlation range is achieved on 15000 meters. This range value is used as a radius of the sector for searching neighbors.

\subsubsection{Result of simulation for structural measurements}
\label{sec:fracan}

\begin{figure}[H]
  \begin{subfigure}{.5\textwidth}
  		\centering
  		\includegraphics[width=.8\linewidth]{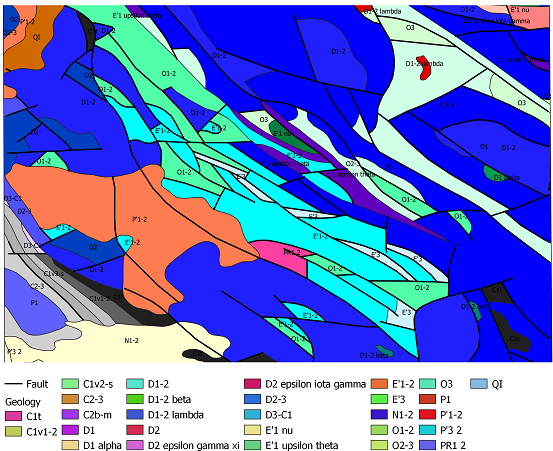}
  		\caption{The image has faults and geology.}
  	\end{subfigure}
  	\begin{subfigure}{.5\textwidth}
  		\centering
  		\includegraphics[width=.8\linewidth]{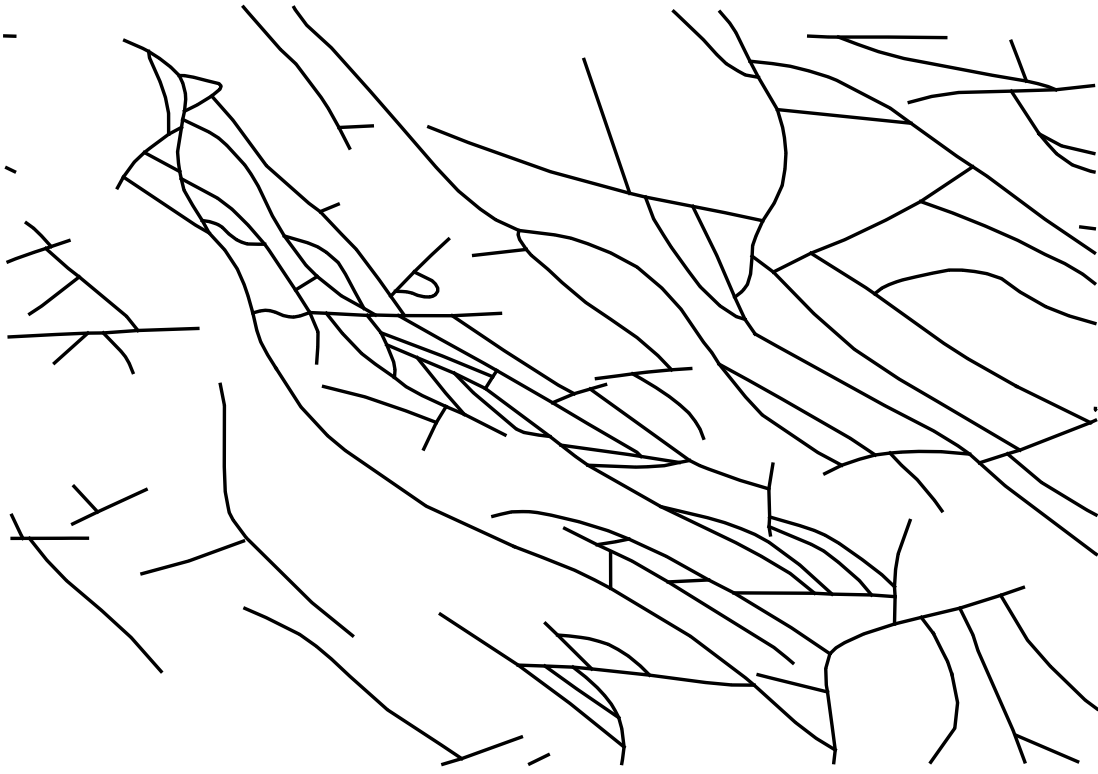}
  		\caption{The image has only faults.}
  	\end{subfigure}
  \caption{Simplified geological.}
  \label{fig:fig10}
\end{figure}

We examined the SGS algorithm in 7011,9 $km^2$ area near of Zhezkazgan city, Kazakhstan. The study area is located in the central part of Kazakhstan. Red beds and volcanic suites are well captured. The rocks wereformed  in age from the Paleozoic, Carboniferous, Devonain and Ordovician. They have a good visibility in defined fold structures. This geological data is obtained from \cite{syusyura2010spatial}. The main purpose of such maps is the investigatigation of  the potential mines in the area, see Figure \ref{fig:fig10}. We take the digital data: geology and faults for the area of interest to make the test of fracture characterization. More detail about the description and color of the legend is explained in source \cite{syusyura2010spatial}.

Based on data analysis from the tool we get next information that is necessary for SGS algorithm:

\begin{itemize}
\item Mean of segment length – 2289.27 meters, The value is used as constant length of simulated fractures segment. As previously noted the length can be specified by an user;
\item Exist bimodal normal distribution of angle faults, 120 and about 300 degrees. 300 degrees direction is the opposite direction (120 degrees) so we choose 120 degrees as the mean of the azimuth angle of distribution for the known domain;
\item Base on the variogram model of segment length, we define lag of between segments - 15000 m. Variogram model is represented in spherical mode.
\end{itemize}

\begin{figure}[H]
  \begin{subfigure}{.5\textwidth}
  		\centering
  		\includegraphics[width=.8\linewidth]{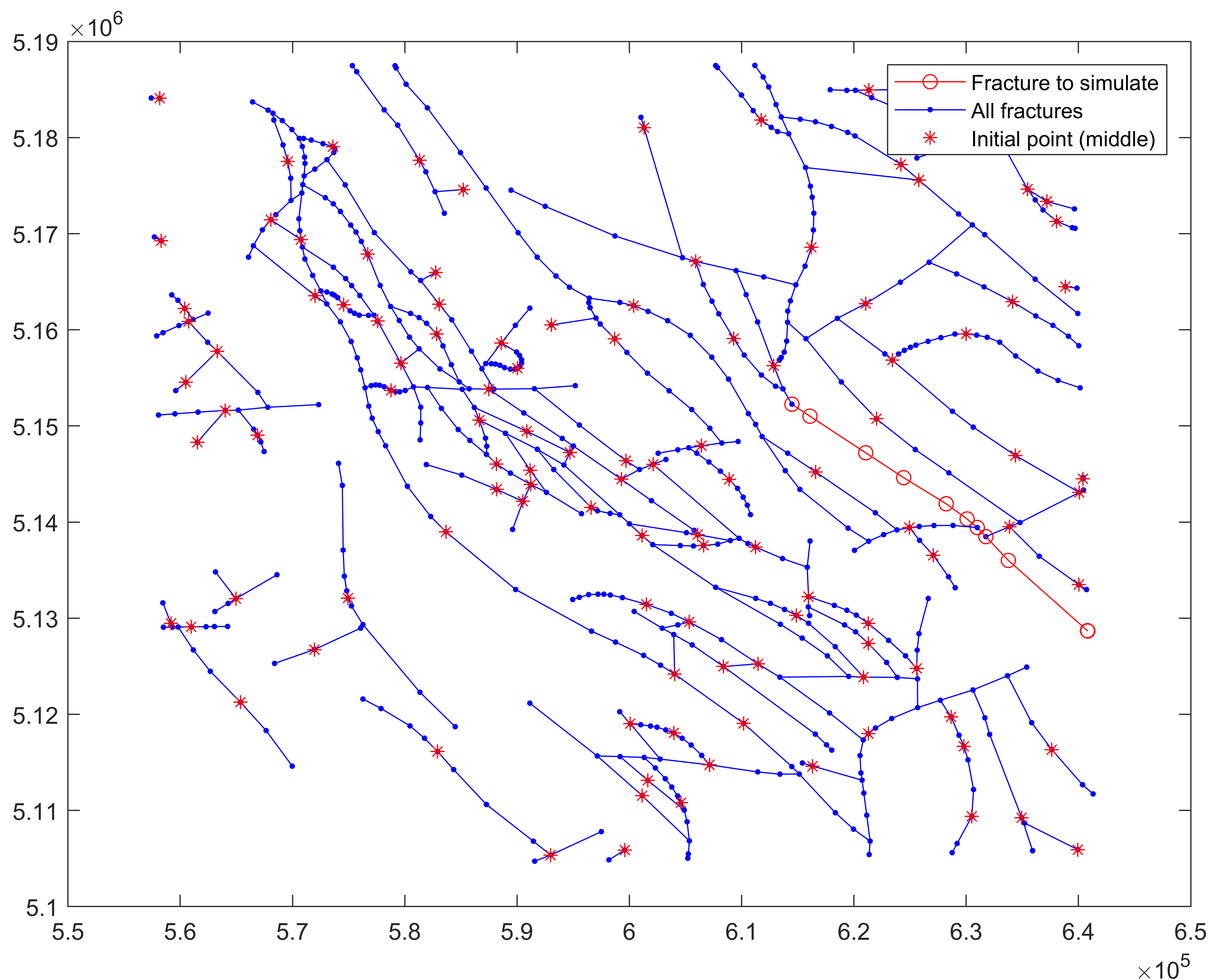}
  		\caption{Geology faults with middle points.}
  		\label{fig:fig11a}
  	\end{subfigure}
  	\begin{subfigure}{.5\textwidth}
  		\centering
  		\includegraphics[width=.8\linewidth]{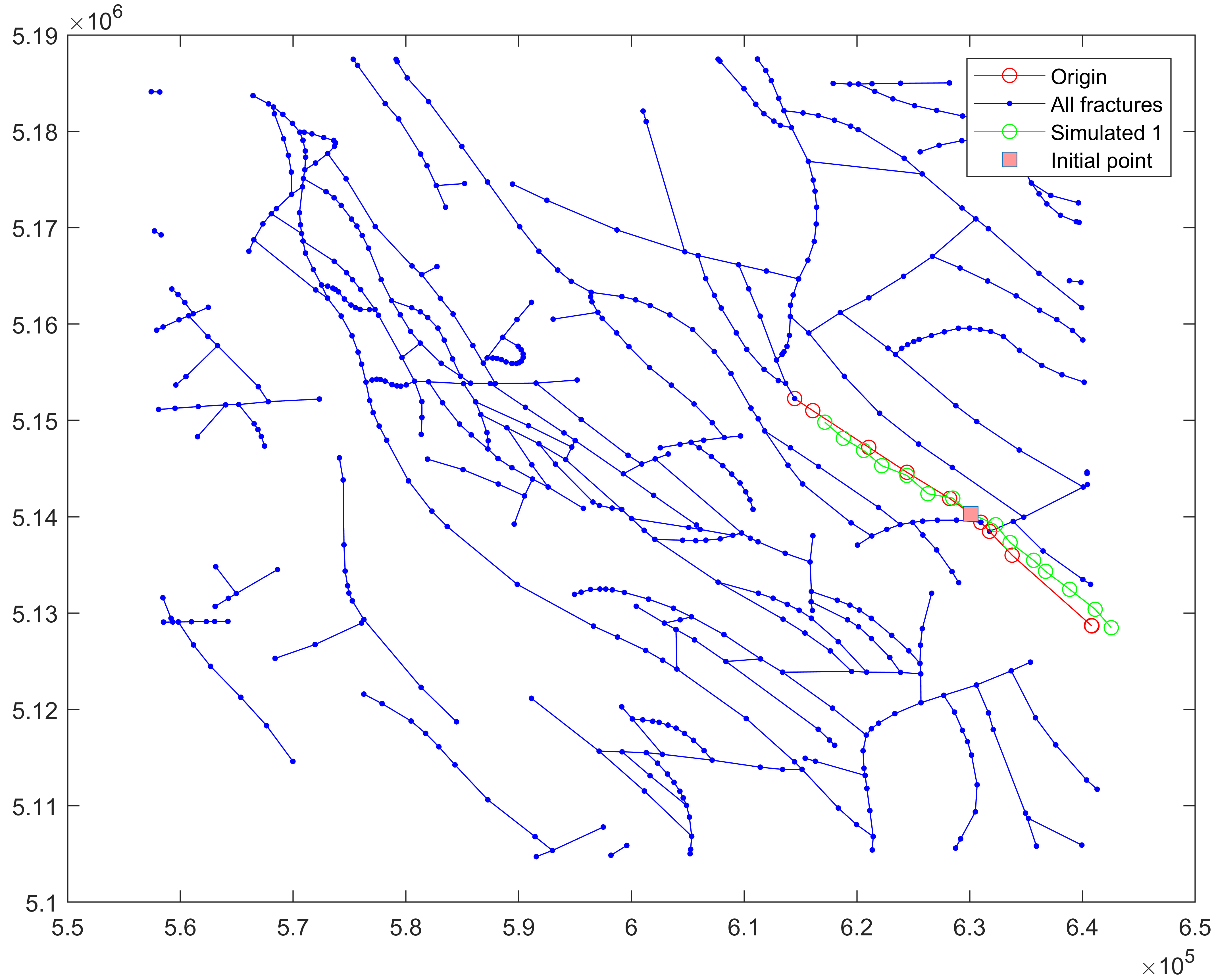}
  		\caption{The simulated chosen fault.}
  		\label{fig:fig11b}
  	\end{subfigure}
  \caption{Result of SGS simulation for one fracture.}
  \label{fig:fig11}
\end{figure}

Propagations for each fracture started from the middle of fracture length and propagate to both sides from a middle point. As shown in Figure \ref{fig:fig11a}, a trace indicated in red was selected to be predicted by the proposed method. After removal of that fracture from the conditing data set, the method used fractures in blue as conditioned data and then a fracture marked in green is generated at the location of the removed fracture. As can be seen from Figure \ref{fig:fig11b}, there is a good match between the original fracture (red) and the simulated fracture(green). 

\begin{figure}[H]
	\begin{subfigure}{.32\textwidth}
		\centering
		\includegraphics[width=1.0\columnwidth]{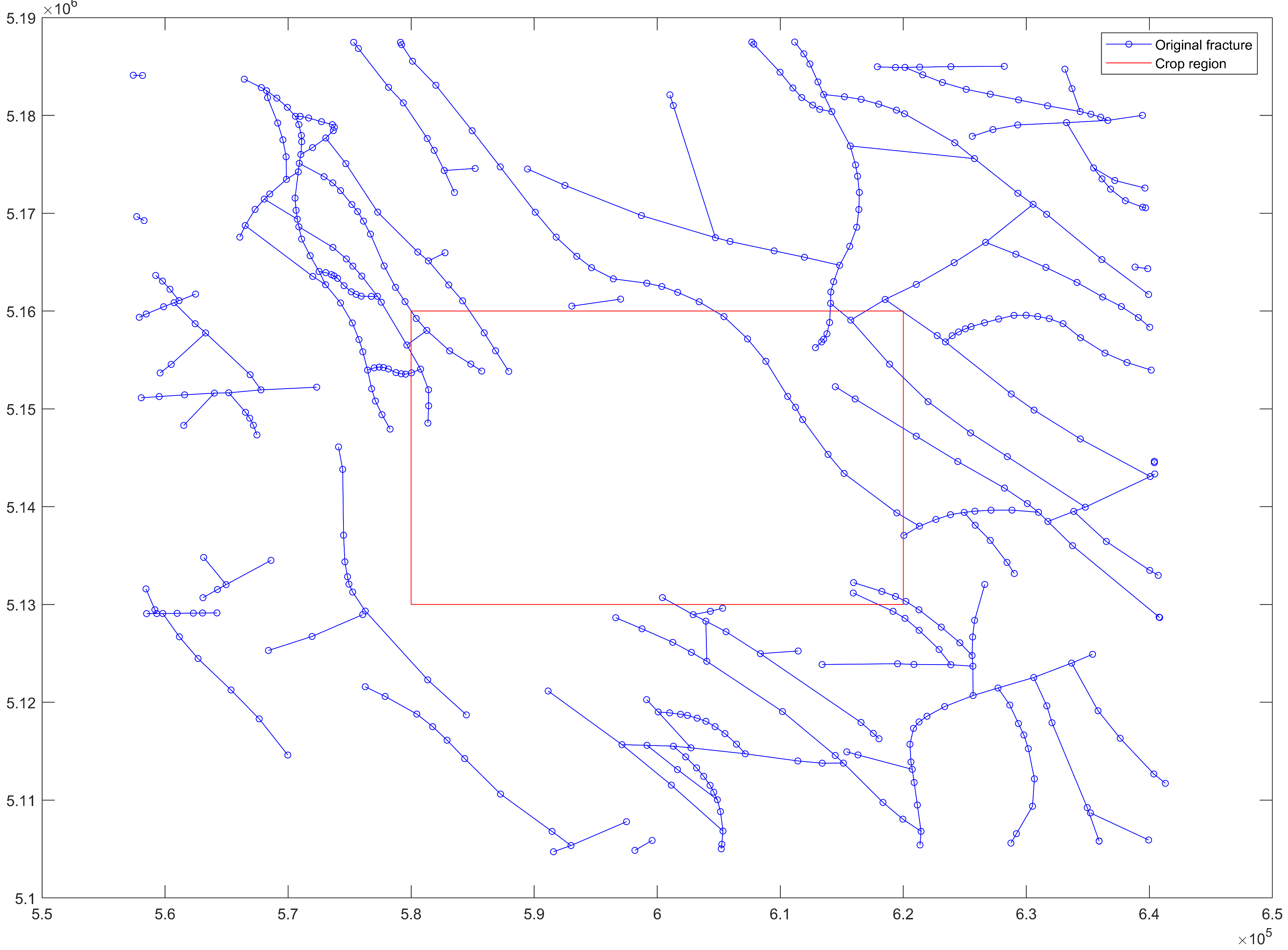}
		\caption{Hidden zone of fractures}
		\label{fig:fig12a}
	\end{subfigure}
	\hfill
	\begin{subfigure}{.32\textwidth}
		\centering  		
		\includegraphics[width=1.0\columnwidth]{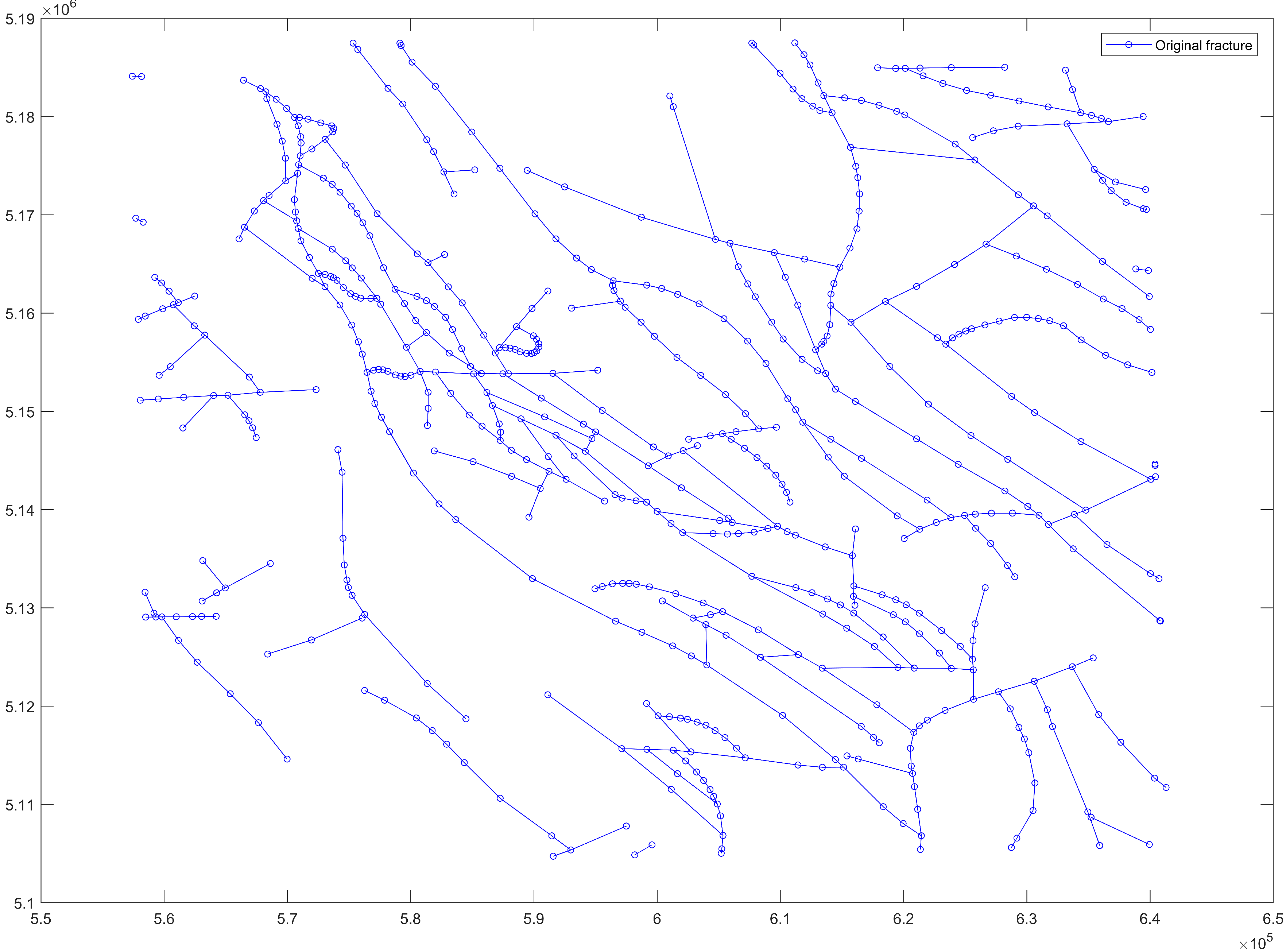}%
		\caption{Original fractures}
		\label{fig:fig12b}
	\end{subfigure}
	\hfill
	\begin{subfigure}{.32\textwidth}
		\centering 
		\includegraphics[width=1.0\columnwidth]{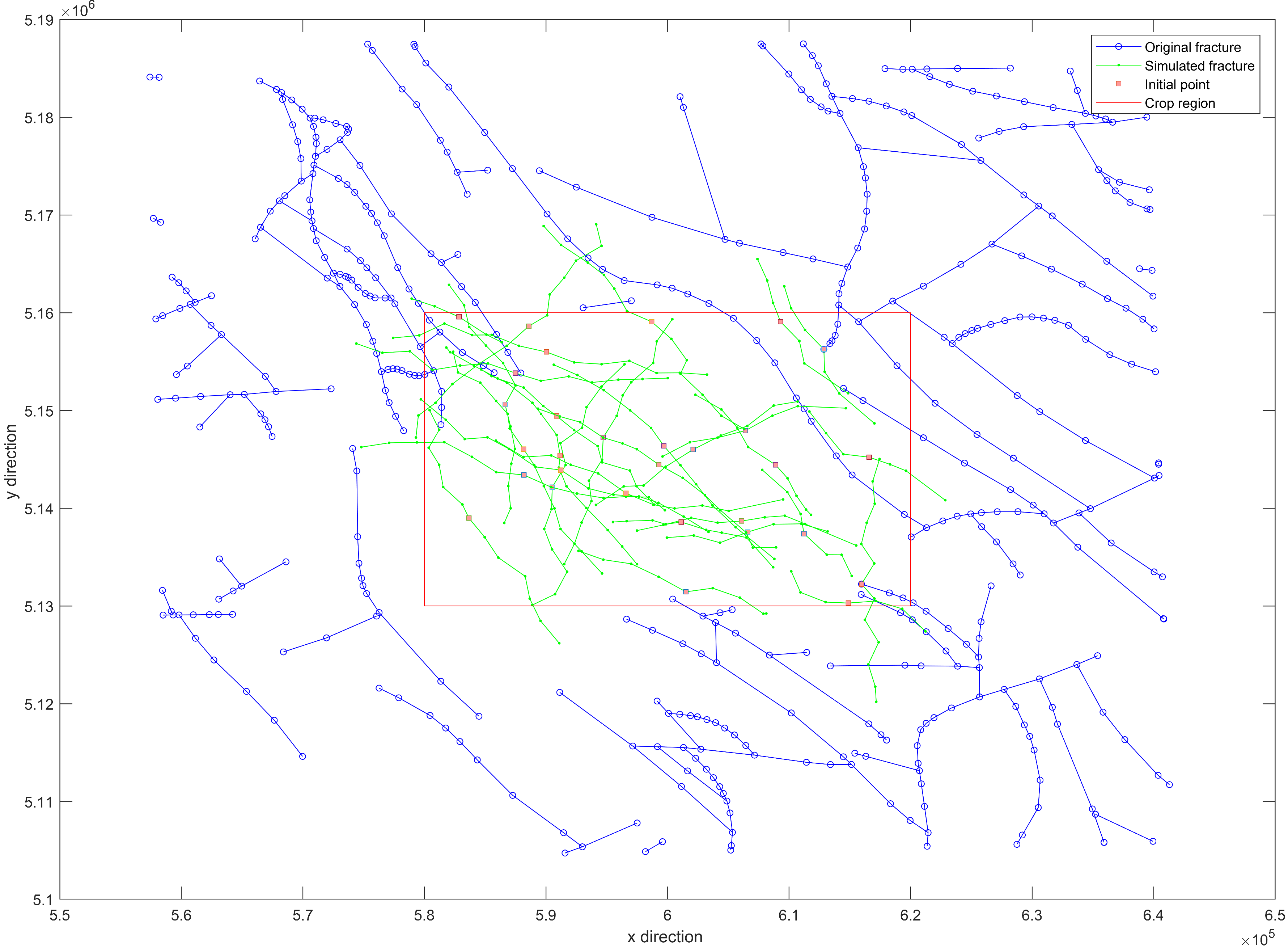}%
		\caption{Simulated fractures}
		\label{fig:fig12c}
	\end{subfigure}
	\caption{Result of Proposed algorithm, simulation for KZ geology fractures.}
	\label{fig:fig12}
\end{figure}

To test the algorithm for realistic condition, we hide the fracture information in the center area (cropped from the original domain). We have used the midpoints of the original fractures as initial seeds for simulation of green fractures. In Figure \ref{fig:fig12a}, red color is the boundary of crop domain from the original fractures, the blue color is original fractures or fractures from the known domain. In Figure \ref{fig:fig12b} the original fractures are in the blue color. In Figure \ref{fig:fig12c} there are the simulated fractures by the proposed algorithm in the green color. The result of the algorithm is consistent with the major trends of the original fracture traces.

Geological fractures are divided into primary and secondary features, The primary fractures are long extensional features, the secondary fractures are of insignificant length, and their prevails over the primary. These fractures have different structures and origins. If the midpoint of long fractures are not inside the hidden square we keep such faults as conditioned data. This is to ensure that we reproduce the length distribution accurately. 

\subsection{Example 3}
\label{sec:expl3}

\begin{figure}[H]
  	\begin{subfigure}{.32\textwidth}
  		\centering
  		\includegraphics[width=1.0\columnwidth]{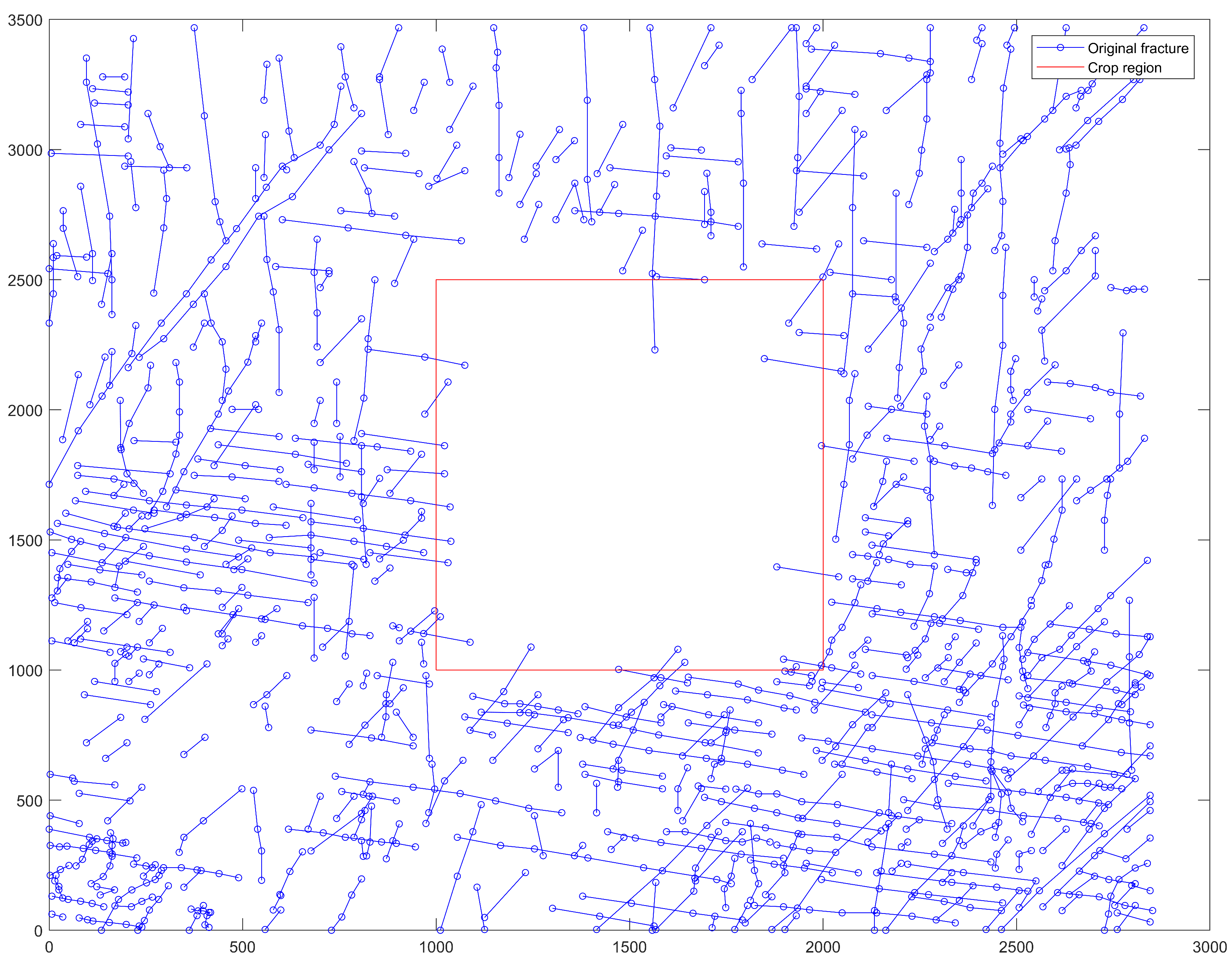}
  		\caption{Hidden zone of fractures}
	\end{subfigure}
\hfill
	\begin{subfigure}{.32\textwidth}
		\centering  		
  		\includegraphics[width=1.0\columnwidth]{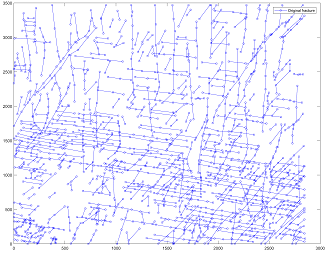}%
		\caption{Original fractures}
	\end{subfigure}
\hfill
	\begin{subfigure}{.32\textwidth}
		\centering 
  		\includegraphics[width=1.0\columnwidth]{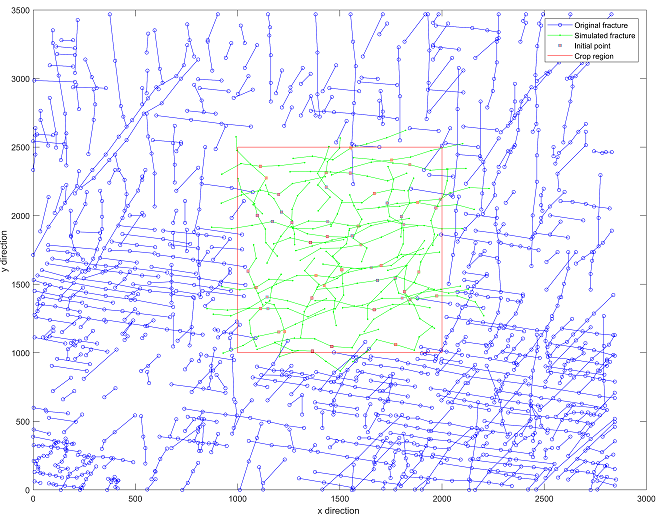}%
  		\caption{Simulated fractures}
	\end{subfigure}
  \caption{Result of proposed algorithm, simulation for McDonald Limestone dataset from the Spireslack open cast coal pit, south of Glasgow in Scotland, UK.}
  \label{fig:fig13}
\end{figure}

The realistic fracture obtained from \cite{healy2017fracpaq} is used in the fracture characterization to illustrate the proposed model. Collected data is a fractured bedding plane in the McDonald (or Hosie) Limestone in the Spireslack opencast coal pit. Overall, in Midland Valley there are mining carboniferous coal-bearing fluvial-deltaic rocks and have been started since the 19th century, Figure \ref{fig:fig13}. The count of initial segment fractures is user-defined for an unknown area. In this case, we choose the 50 initial segment fractures. 50 initial fracture segments randomly seed by the Poisson process in the hidden center area. 

As the previous figure, we got 3 images: hidden domain (cropped region), original fractures and simulated in Figure \ref{fig:fig13}. If we compare the result of fractures simulation and original fractures on the figures, we can see that the trend of fractures direction is reproduced. To prove this statement we will resort to a histogram of fracture angles distribution.

\begin{figure}[H]
	\begin{tabular}{cc}
  \begin{subfigure}{.5\textwidth}
  		\centering
  		\includegraphics[width=.8\columnwidth]{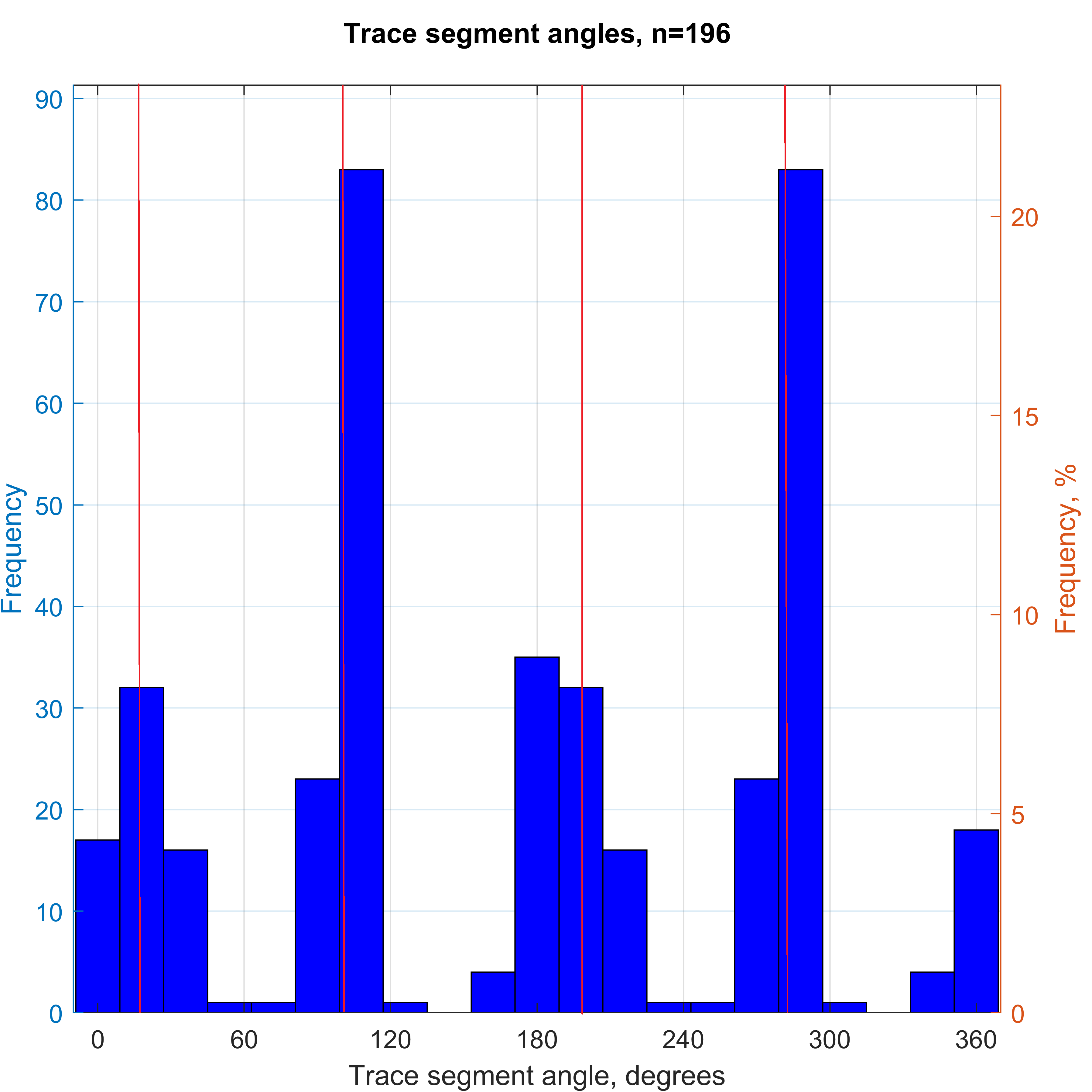}
  		\caption{For original fractures.}
  	\end{subfigure}
  	\begin{subfigure}{.5\textwidth}
  		\centering
  		\includegraphics[width=.8\columnwidth]{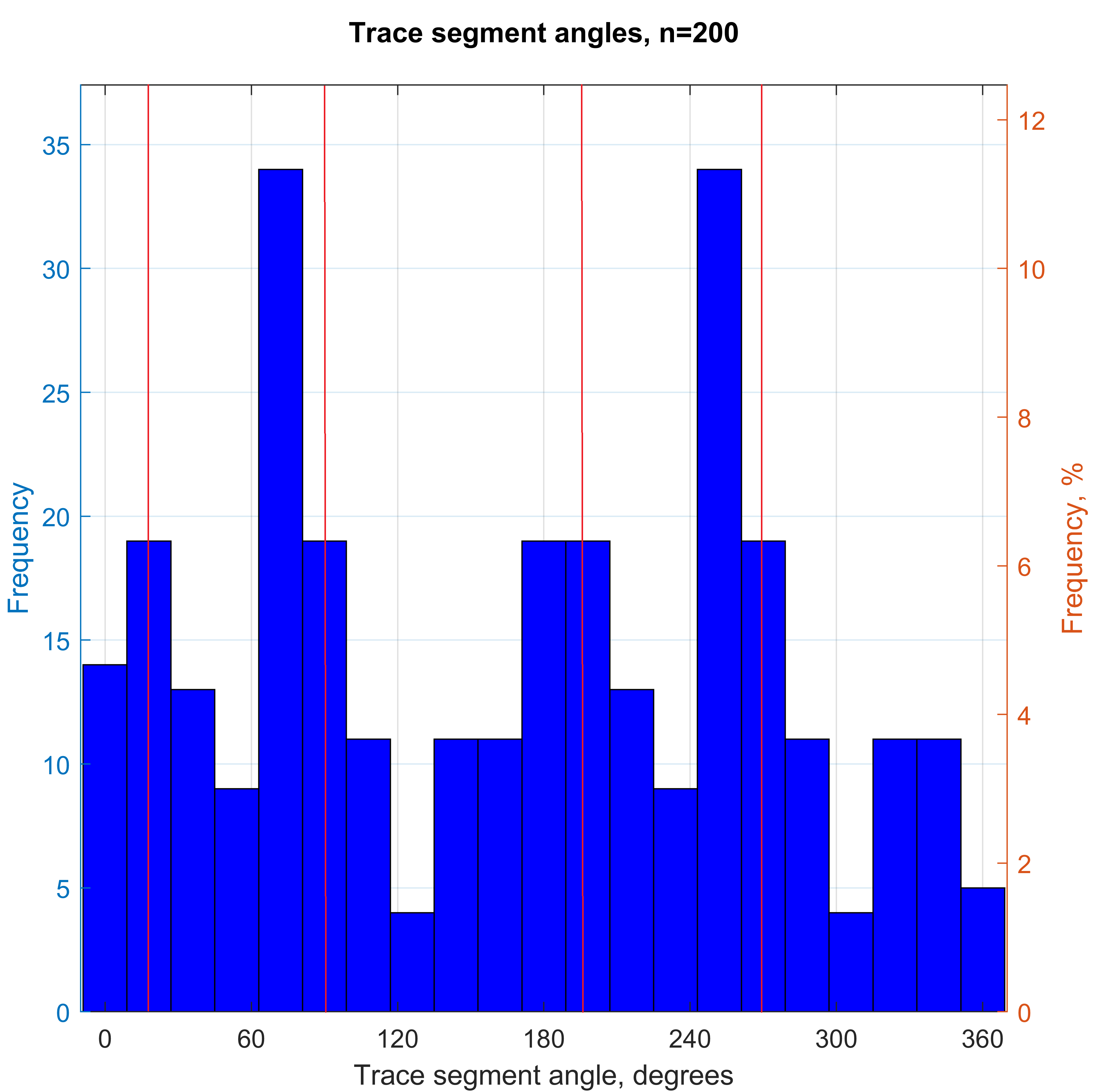}
  		\caption{For simulated fractures.}
  	\end{subfigure}
  	\end{tabular}
  \caption{Histograms of verification.}
  \label{fig:fig14}
\end{figure}

The verification was given by comparing the distribution of fracture angles for original fracture networks (left figure) and simulated fracture networks (right figure) within the hidden area, see Figure \ref{fig:fig14}.  We have observed similarity between histograms for original and simulated fractures. In particular, shape and means of distribution are similar that was illustrated in Figure \ref{fig:fig14} (vertical red lines in histogram).

Analysis of the plotted histograms yielded a multimodal distribution (4 peaks), it is due to the fact that we have two main directions of fractures and two conjugate directions for propagation of fractures. Based on histogram we get means of trace segment angles for original fracture networks: 20; 110; 185; 290. The corresponding angles of the simulated fracture networks: 18; 85; 190; 275. These histograms give us the opportunity to assert that the proposed method shows a good result and confirms its ability to simulate fracture networks.

\begin{figure}[H]
	\centering
	\includegraphics[scale=3.0]{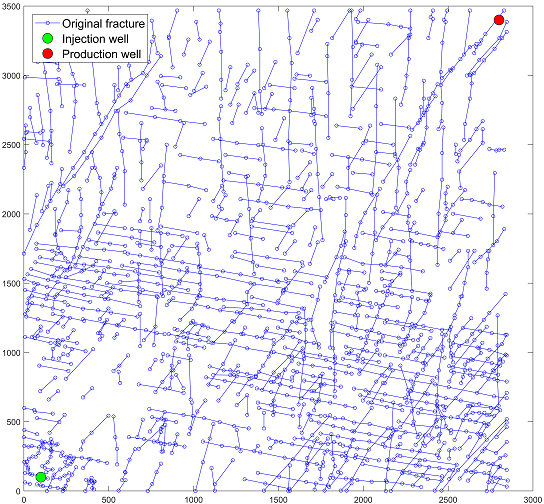}
	\caption{Location of injection and production wells.}
	\label{fig:fig15}
\end{figure}

 The additional verification of the fracture network model is given by comparing the concentration profile at the production well in a tracer test setting. We have a single-phase , incompressible flow and transport model which is based on work presented in \cite{amanbek2019adaptive}. Here, the flow problem solved initially and then the transport problem. For more details we refer to earlier work \cite{amanbek2019adaptive, singh2017adaptive, amanbek2019recovery, amanbek2018new, thomas2011enhanced, amanbek2020error}. We locate the production well in the top right corner of the domain and the injection well in the bottom left of the domain, see Figure \ref{fig:fig15}.   Dimension of the reservoir is 6900 ft x 5700 ft. In the adaptive regime, the local cell dimension is 10 x 10 grids for homogenization of permeability, i.e. domain dimension is 69 x 57 grids in coarse scale. Porosity was taken as 0.2 and diffusion is 0.0001 $ft^2/day$. No flow boundary conditions in the domain. Permeability at the fracture was chosen 200 times more than non-fracture media. Time step is 5 days, ratio between coarse and fine scale is 10 and dx=10 ft. The injection rate at the bottom left well is chosen 100 stb/day. 

\begin{figure}[H]
  \centering
  \includegraphics[scale=0.25]{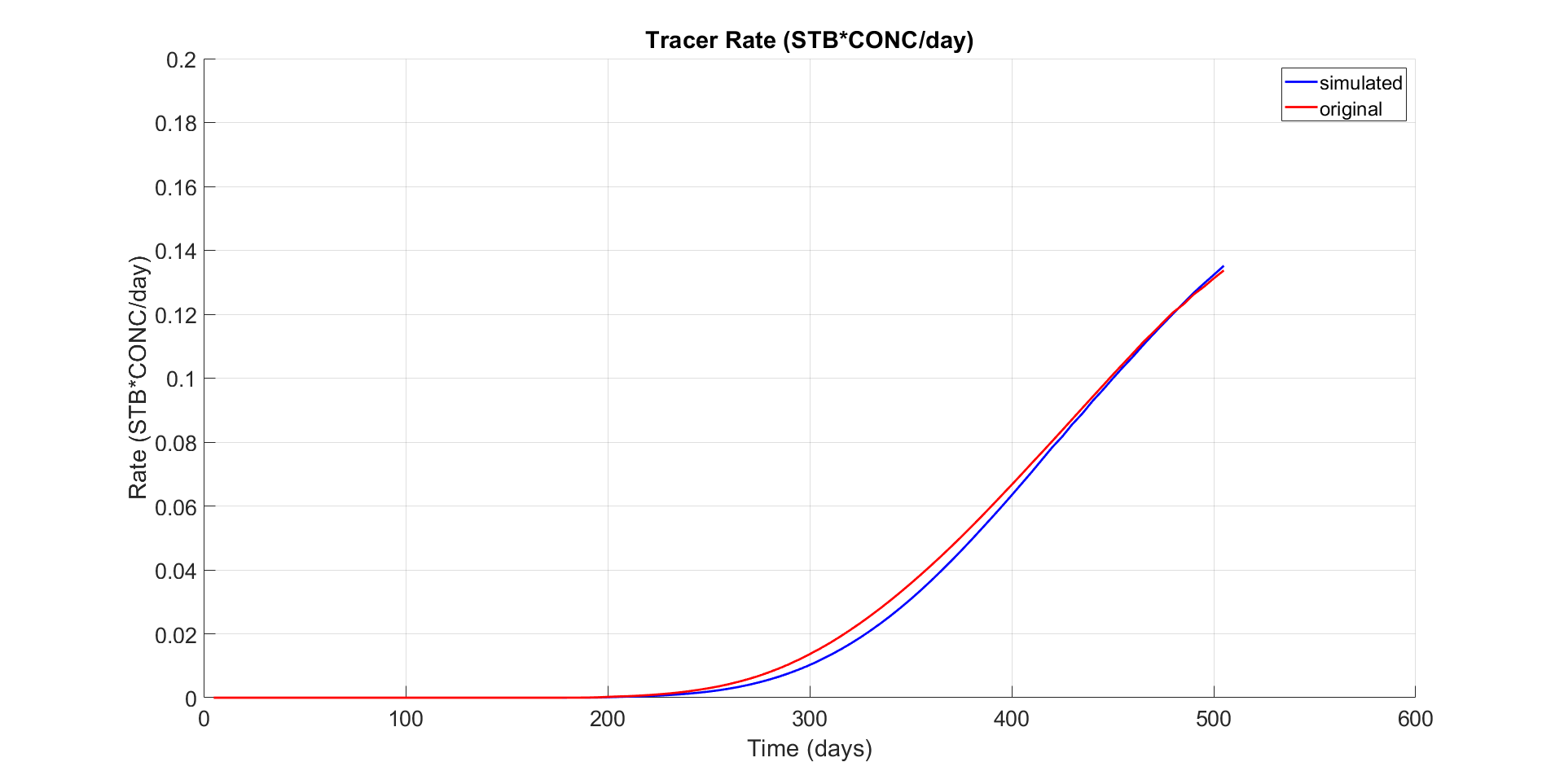}
  \caption{Location of injection and production wells.}
  \label{fig:fig16}
\end{figure}

As shown in Figure \ref{fig:fig16}, the flow and transport model with proposed fracture network algorithm was able to match concentration at production well for the model with original fracture network. There is a good agreement in concentration profile for both scenarios as well as on breakthrough time. We limit in this work with provided realization of fracture network in Figure \ref{fig:fig14} to illustrate the verification. 

Our verification was performed for the provided realization. Future work will involve more flow and transport simulations for various realizations in an efficient way using the multicore machines in parallel mode.

\section{Discussion}
\label{sec:disc}
One can look at the history of fracture generation in porous media. In other words, if model allows to go back in time and afterwards to take into account main geological events from the past which was reason for generation of younger fractures along with older ones. Thus, our method can be seen as similar as the fracture propagation scenario where it starts from seed nodes obtained from the Poisson Distribution or  user-specified locations. In addition,  the algorithm can be conditioned by data from available sources such as seismic, wellbore and others. 

We assumed that the fracture orientation has a gaussian distribution in the simulation regime.  The understanding of behaviour of 2D fracture network is an important base for studying the fracture network in 3D. The extension from two dimensional to three dimensional can be done in a similar fashion by considering azimuth, length and polar angles. A combination of layers can be interpreted as characterization of the subsurface volume \cite{srivastava2005geostatistical}. During the generation of a sector zone, we have used the unique and precomputed standard deviation and radius for each sector in the simulation process. However, these parameters can be specified by user. 

\section{Conclusion}
\label{sec:concl}
The purpose of this paper is to  provide the framework of the characterization of fracture network in different scales using the geostatistical method. We have applied the proposed algorithm for natural fracture path  from  the Central Kazakhstan, the area includes Zhezkazgan city, and for McDonald limestone from Scotland, UK. To our best knowledge, this is the first study to deal with the fracture network modeling of the Central Kazakhstan faults. In the growth-based fracture network model, the Gaussian Sequential Simulation is used for orientation of fracture honoring the neighborhood information. We currently believe that the computation time of the fracture network characterization using the proposed algorithm is less than the CPU time of the geomechanical models that considers the coupled partial differential equations in the subsurface. To verify the proposed fracture network method, the distribution of the segment angles for the original fracture path is compared with the same distribution for the simulated fracture path. The main features of both distributions are in good agreement. In addition, we have compared the concentration profiles at the production well for different fracture networks in the single-phase and incompressible flow and transport models. There is a good match between the concentration history used simulated fracture network and the concentration history used original fracture network. 
In our future research we intend to concentrate on machine learning algorithms for fracture characterization using image training approaches for real-field dataset. 

\section*{Acknowledgement}
\label{sec:ackn}

Financial support from the Nazarbayev University Faculty Development Competitive Research Grant,  No 110119FD4502, is gratefully acknowledged.

\bibliographystyle{unsrt}  
\bibliography{FNM_CondData_SGS}  

%
%
%
%

\end{document}